\documentclass[reprint,superscriptaddress]{revtex4-1}

\usepackage{color}
\usepackage{graphicx}
\usepackage{hyperref}
\usepackage{amsmath}
\usepackage{array}
\usepackage{subcaption}
\usepackage{float}
\usepackage[justification=raggedright]{caption}
\begin{document}

\title{Dark Matter Results from First 98.7-day Data of PandaX-II Experiment}
\date{\today}
\affiliation{INPAC and Department of Physics and Astronomy, Shanghai Jiao Tong University, Shanghai Laboratory for Particle Physics and Cosmology, Shanghai 200240, China}
\author{Andi Tan}
\affiliation{Department of Physics, University of Maryland, College Park, Maryland 20742, USA}
\author{Mengjiao Xiao}
\author{Xiangyi Cui}
\author{Xun Chen}
\affiliation{INPAC and Department of Physics and Astronomy, Shanghai Jiao Tong University, Shanghai Laboratory for Particle Physics and Cosmology, Shanghai 200240, China}
\author{Yunhua Chen}
\affiliation{Yalong River Hydropower Development Company, Ltd., 288 Shuanglin Road, Chengdu 610051, China}
\author{Deqing Fang}
\affiliation{Shanghai Institute of Applied Physics, Chinese Academy of Sciences, 201800, Shanghai, China}
\author{Changbo Fu}
\author{Karl Giboni}
\affiliation{INPAC and Department of Physics and Astronomy, Shanghai Jiao Tong University, Shanghai Laboratory for Particle Physics and Cosmology, Shanghai 200240, China}
\author{Franco Giuliani}
\affiliation{INPAC and Department of Physics and Astronomy, Shanghai Jiao Tong University, Shanghai Laboratory for Particle Physics and Cosmology, Shanghai 200240, China}
\author{Haowei Gong}
\author{Ke Han}
\affiliation{INPAC and Department of Physics and Astronomy, Shanghai Jiao Tong University, Shanghai Laboratory for Particle Physics and Cosmology, Shanghai 200240, China}
\author{Shouyang Hu}
\affiliation{Key Laboratory of Nuclear Data, China Institute of Atomic Energy, Beijing 102413, China}
\author{Xingtao Huang}
\affiliation{School of Physics and Key Laboratory of Particle Physics and Particle Irradiation (MOE), Shandong University, Jinan 250100, China}
\author{Xiangdong Ji}
\email[Spokesperson: ]{xdji@sjtu.edu.cn}
\affiliation{INPAC and Department of Physics and Astronomy, Shanghai Jiao Tong University, Shanghai Laboratory for Particle Physics and Cosmology, Shanghai 200240, China}
\affiliation{Center of High Energy Physics, Peking University, Beijing 100871, China}
\affiliation{Department of Physics, University of Maryland, College Park, Maryland 20742, USA}
\author{Yonglin Ju}
\affiliation{School of Mechanical Engineering, Shanghai Jiao Tong University, Shanghai 200240, China}

\author{Siao Lei}
\author{Shaoli Li}
\affiliation{INPAC and Department of Physics and Astronomy, Shanghai Jiao Tong University, Shanghai Laboratory for Particle Physics and Cosmology, Shanghai 200240, China}

\author{Xiaomei Li}
\affiliation{Key Laboratory of Nuclear Data, China Institute of Atomic Energy, Beijing 102413, China}

\author{Xinglong Li}
\affiliation{Key Laboratory of Nuclear Data, China Institute of Atomic Energy, Beijing 102413, China}

\author{Hao Liang}
\affiliation{Key Laboratory of Nuclear Data, China Institute of Atomic Energy, Beijing 102413, China}

\author{Qing Lin}
\thanks{Now at Department of Physics, Columbia University}
\affiliation{INPAC and Department of Physics and Astronomy, Shanghai Jiao Tong University, Shanghai Laboratory for Particle Physics and Cosmology, Shanghai 200240, China}


\author{Huaxuan Liu}
\affiliation{School of Mechanical Engineering, Shanghai Jiao Tong University, Shanghai 200240, China}

\author{Jianglai Liu}
\email[Corresponding author: ]{jianglai.liu@sjtu.edu.cn}
\affiliation{INPAC and Department of Physics and Astronomy, Shanghai Jiao Tong University, Shanghai Laboratory for Particle Physics and Cosmology, Shanghai 200240, China}

\author{Wolfgang Lorenzon}
\affiliation{Department of Physics, University of Michigan, Ann Arbor, MI, 48109, USA}

\author{Yugang Ma}
\affiliation{Shanghai Institute of Applied Physics, Chinese Academy of Sciences, 201800, Shanghai, China}
\author{Yajun Mao}
\affiliation{School of Physics, Peking University, Beijing 100871, China}

\author{Kaixuan Ni}
\thanks{Now at Department of Physics, University of California, San Diego}
\affiliation{INPAC and Department of Physics and Astronomy, Shanghai Jiao Tong University, Shanghai Laboratory for Particle Physics and Cosmology, Shanghai 200240, China}


\author{Xiangxiang Ren}
\affiliation{INPAC and Department of Physics and Astronomy, Shanghai Jiao Tong University, Shanghai Laboratory for Particle Physics and Cosmology, Shanghai 200240, China}

\author{Michael Schubnell}
\affiliation{Department of Physics, University of Michigan, Ann Arbor, MI, 48109, USA}

\author{Manbin Shen}
\affiliation{Yalong River Hydropower Development Company, Ltd., 288 Shuanglin Road, Chengdu 610051, China}

\author{Fang Shi}
\affiliation{INPAC and Department of Physics and Astronomy, Shanghai Jiao Tong University, Shanghai Laboratory for Particle Physics and Cosmology, Shanghai 200240, China}


\author{Hongwei Wang}
\affiliation{Shanghai Institute of Applied Physics, Chinese Academy of Sciences, 201800, Shanghai, China}

\author{Jiming Wang}
\affiliation{Yalong River Hydropower Development Company, Ltd., 288 Shuanglin Road, Chengdu 610051, China}

\author{Meng Wang}
\affiliation{School of Physics and Key Laboratory of Particle Physics and Particle Irradiation (MOE), Shandong University, Jinan 250100, China}

\author{Qiuhong Wang}
\affiliation{Shanghai Institute of Applied Physics, Chinese Academy of Sciences, 201800, Shanghai, China}

\author{Siguang Wang}
\affiliation{School of Physics, Peking University, Beijing 100871, China}


\author{Xuming Wang}
\affiliation{INPAC and Department of Physics and Astronomy, Shanghai Jiao Tong University, Shanghai Laboratory for Particle Physics and Cosmology, Shanghai 200240, China}

\author{Zhou Wang}
\affiliation{School of Mechanical Engineering, Shanghai Jiao Tong University, Shanghai 200240, China}

\author{Shiyong Wu}
\affiliation{Yalong River Hydropower Development Company, Ltd., 288 Shuanglin Road, Chengdu 610051, China}

\author{Xiang Xiao}
\author{Pengwei Xie}
\email[Corresponding author: ]{willandy@sjtu.edu.cn}
\affiliation{INPAC and Department of Physics and Astronomy, Shanghai Jiao Tong University, Shanghai Laboratory for Particle Physics and Cosmology, Shanghai 200240, China}

\author{Binbin Yan}
\affiliation{School of Physics and Key Laboratory of Particle Physics and Particle Irradiation (MOE), Shandong University, Jinan 250100, China}

\author{Yong Yang}
\affiliation{INPAC and Department of Physics and Astronomy, Shanghai Jiao Tong University, Shanghai Laboratory for Particle Physics and Cosmology, Shanghai 200240, China}

\author{Jianfeng Yue}
\affiliation{Yalong River Hydropower Development Company, Ltd., 288 Shuanglin Road, Chengdu 610051, China}

\author{Xionghui Zeng}
\affiliation{Yalong River Hydropower Development Company, Ltd., 288 Shuanglin Road, Chengdu 610051, China}

\author{Hongguang Zhang}
\affiliation{INPAC and Department of Physics and Astronomy, Shanghai Jiao Tong University, Shanghai Laboratory for Particle Physics and Cosmology, Shanghai 200240, China}

\author{Hua Zhang}
\affiliation{School of Mechanical Engineering, Shanghai Jiao Tong University, Shanghai 200240, China}
\author{Huanqiao Zhang}
\affiliation{Key Laboratory of Nuclear Data, China Institute of Atomic Energy, Beijing 102413, China}

\author{Tao Zhang}
\author{Li Zhao}
\affiliation{INPAC and Department of Physics and Astronomy, Shanghai Jiao Tong University, Shanghai Laboratory for Particle Physics and Cosmology, Shanghai 200240, China}

\author{Jing Zhou}
\affiliation{Key Laboratory of Nuclear Data, China Institute of Atomic Energy, Beijing 102413, China}
\author{Ning Zhou}
\affiliation{INPAC and Department of Physics and Astronomy, Shanghai Jiao Tong University, Shanghai Laboratory for Particle Physics and Cosmology, Shanghai 200240, China}
\affiliation{Department of Physics, Tsinghua University, Beijing 100084, China}
\author{Xiaopeng Zhou}
\affiliation{School of Physics, Peking University, Beijing 100871, China}

\collaboration{PandaX-II Collaboration}
\begin{abstract}
  We report the WIMP dark matter search results using the first
  physics-run data of the PandaX-II 500 kg liquid xenon dual-phase
  time-projection chamber, operating at the China JinPing underground
  Laboratory.  No dark matter candidate is identified above
  background.  In combination with the data set during the
  commissioning run, with a total exposure of 3.3$\times10^4$ kg-day,
  the most stringent limit to the spin-independent interaction between
  the ordinary and WIMP dark matter is set for a range of dark matter
  mass between 5 and 1000 GeV/c$^2$. The best upper limit on the
  scattering cross section is found $2.5\times 10^{-46}$ cm$^2$ for
  the WIMP mass 40 GeV/c$^2$ at 90\% confidence level.

\end{abstract}
\maketitle

Weakly interacting massive particles, WIMPs in short, are a class of
hypothetical particles that came into existence shortly after the Big
Bang. The WIMPs could naturally explain the astronomical and
cosmological evidences of dark matter in the Universe.  The weak
interactions between WIMPs and ordinary matter could lead to the
recoils of atomic nuclei that produce detectable signals in
deep-underground direct detection experiments. Over the past decade,
the dual-phase xenon time-projection chambers (TPC) emerged as a
powerful technology for WIMP searches both in scaling up the target
mass, as well as in improving background
rejection~\cite{Aprile:2006kx,Aprile:2012nq,Akimov:2011tj}. LUX, a
dark matter search experiment with a 250 kg liquid xenon target, has
recently reported the best limit of 6$\times 10^{-46}$~cm$^2$ on the
WIMP-nucleon scattering cross section~\cite{Akerib:2015rjg}, with no
positive signals observed. The PandaX-II experiment, a half-ton scale
dual-phase xenon experiment at the China JinPing underground
Laboratory (CJPL), has recently reported the dark matter search
results from its commissioning run (Run 8, 19.1 live days) with a
5845~kg-day exposure~\cite{Tan:2016diz}. The data were contaminated with significant
$^{85}$Kr background.  After a krypton distillation campaign in early
2016, PandaX-II commenced physics data taking in March 2016. In
this paper, we report the combined WIMP search results using the data
from the first physics run from March 9 to June 30, 2016 (Run 9, 79.6
live days) and Run 8, with a total of 3.3$\times$10$^4$ kg-day
exposure, the largest reported WIMP data set among dual-phase xenon
detectors in the world to date.

The PandaX-II detector has been described in detail in
Ref.~\cite{Tan:2016diz}. The liquid xenon target consists of a
cylindrical TPC with dodecagonal cross section (opposite-side
distance 646~mm), confined by the polytetrafluoroethylene (PTFE)
reflective wall, and a vertical drift distance of 600~mm defined by the 
cathode mesh and gate grid located at the bottom and top.
For each physical event, the prompt scintillation photons (S1) and the delayed
electroluminescence photons (S2) from the ionized electrons are
collected by two arrays of 55 Hamamatsu R11410-20 photomultiplier
tubes (PMTs) located at the top and bottom, respectively. 
This allows reconstruction of
an event energy and position. A skin liquid xenon region outside of
the PTFE wall was instrumented with 48 Hamamatsu R8520-406 1-inch PMTs
serving as an active veto. The $\gamma$ background, which produces
electron recoil (ER) events, can be distinguished from the dark matter
nuclear recoil (NR) using the S2-to-S1 ratio.

During the data taking period in Run 9, a few different TPC field 
settings were used at different running periods (Table~\ref{tab:run_sets}) 
to maximize the drift and electron extraction fields while avoiding spurious 
photons and electrons emission from the electrodes.
In each period, calibration runs were taken
to study the detector responses.  The PMT gains were calibrated using
the single photoelectrons (PEs) produced by the low-intensity
blue-light pulses transmitted into the detector. To study the multiple
PE production in R11410-20 by the approximately 178~nm photons in
xenon~\cite{Faham:2015kqa}, we measured the PE distributions for
individual PMTs using low intensity ($<$3 PE) physical S1 signals.  An
average double PE fraction (DPF, defined as the ratio between the 
occurrence of double PE emission to the total occurrence of non-zero PEs) of
0.21$\pm$0.02 was obtained. The position-dependent responses
in S1 and S2 were calibrated using the 164 keV $\gamma$
peak from meta-stable $^{131\rm{m}}$Xe, produced by exposure 
to cosmic rays at ground and neutrons during NR calibration. 
Towards the end of Run~9, the electron lifetime reached 940$\pm$50 $\mu$s, compared to
the maximum electron drift time 350~$\mu$s in the TPC. 
The non-uniformities for S1 (3D)
and S2 (horizontal plane) were both measured to be less than 10\% in standard 
deviation in the fiducial volume (FV, defined later).
The photon detection efficiency (PDE), the electron extraction efficiency (EEE),
and the single electron gain (SEG) and their uncertainties are also
shown in Table~\ref{tab:run_sets}.

\begin{table}
  \begin{tabular}{cccccccc}
    \hline\hline
    Setting & live time & $E_{\rm drift}$ & $E_{\rm extract}$ & PDE & EEE & SEG & $\tau_{e}$\\
    & (day) & (V/cm) & (kV/cm) & (\%) & (\%) & PE/e & ($\mu$s)\\\hline
    1 & 7.76 & 397.3 & 4.56 & 11.76 & 46.04 & 24.4 & 348.2  \\
    2 & 6.82 & 394.3 & 4.86 & 11.76 & 54.43 & 26.9 & 393.1 \\
    3 & 1.17 & 391.9 & 5.01 & 11.76 & 59.78 & 26.7 & 409.0 \\
    4 & 63.85 & 399.3 & 4.56 & 11.76 & 46.04 & 24.4 & 679.6 \\
    \hline\hline
  \end{tabular}
  \caption{Summary of four settings in Run 9 where
    $E_{\rm drift}$ and $E_{\rm extract}$ are the drift field in the liquid
    and electron extraction field in the gaseous xenon, respectively.
    The field values were
    calculated based on a COMSOL simulation. 
    The fractional uncertainties of the PDE, EEE and
    SEG are 5\%, 6\% and 3\%, respectively, for all setting.
    Only the average electron lifetime is given in the table, although the S2 vertical
    uniformity correction was performed for every data taking unit, 
    typically lasting for
    24 hours, based on the electron lifetime obtained therein.
  }
  \label{tab:run_sets}
\end{table}

The analysis reported in this paper follows the procedure as in
Ref.~\cite{Tan:2016diz} with the following major differences. An
improved position reconstruction was developed based on a data-driven
photon acceptance function (PAF) similar to Ref.~\cite{6359814},
leading to a better FV definition. We performed NR
calibration runs using a low-intensity (approximately 2 Hz) $^{241}$Am-Be
(AmBe) neutron source with improved statistics, and an ER calibration
run by injecting tritiated methane~\cite{Akerib:2015wdi}, leading to a
better understanding of the distributions of the NR and ER events.  A
boosted-decision-tree (BDT) cut method was developed to suppress
accidental background by more than a factor of three. Together with a
much larger exposure and lower Kr background, these improvements
account for the much more sensitive dark matter search results reported
here.

The data quality cuts used in this analysis follow those in
Refs.~\cite{Xiao:2015psa} and~\cite{Tan:2016diz}, 
including noise filters on ripple-like waveforms, cuts on S1 and S2 to 
remove events with abnormal PMT pattern or top/bottom ratio, 
and a 3-PMT coincidence and no-more-than two ``S1-like'' signals 
per event requirement to suppress random coincidence.

Most of external background events are located close to the detector boundary, 
therefore a powerful background rejection demands good position
reconstruction. In addition to the template matching (TM) algorithm in
Ref.~\cite{Xiao:2015psa}, a new algorithm was developed based on an
iterative fitting of the position-dependent PAF for each PMT. For PMTs
located close to the PTFE wall, the PAFs took into account effects due
to photon reflections.  In each iteration, the position was
reconstructed by maximizing the charge likelihood according to the PAF
obtained from the previous iteration, and the new position entered
into the determination of the next PAF. Using this reconstruction, the
Kr events in Run 8 and the tritium calibration events in Run 9 yielded
good uniformity in the horizontal plane.

\begin{figure}[h]
  \centering
  \begin{subfigure}[b]{.45\textwidth}
    \includegraphics[width=\textwidth]{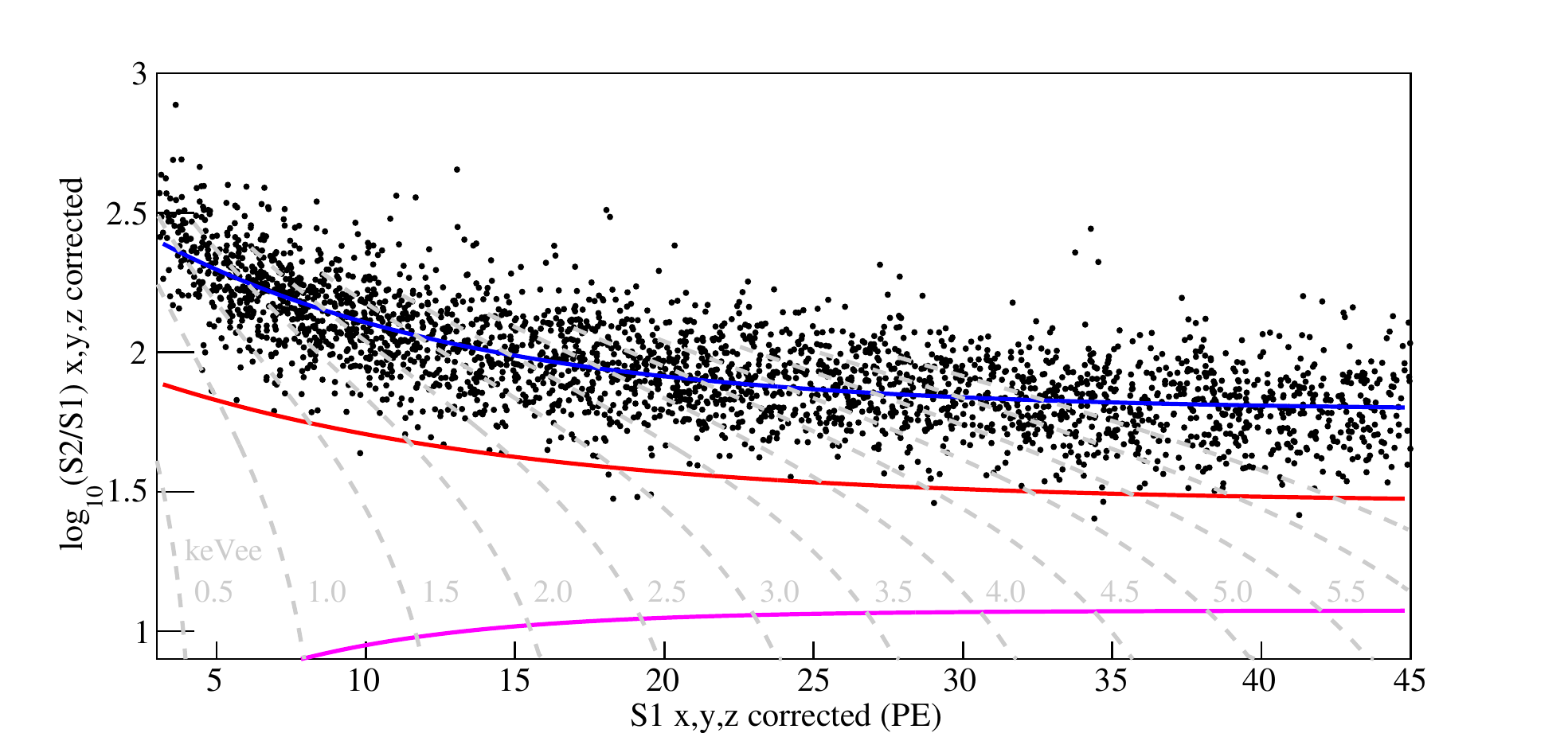}
  \end{subfigure}
  \begin{subfigure}[b]{.45\textwidth}
    \includegraphics[width=\textwidth]{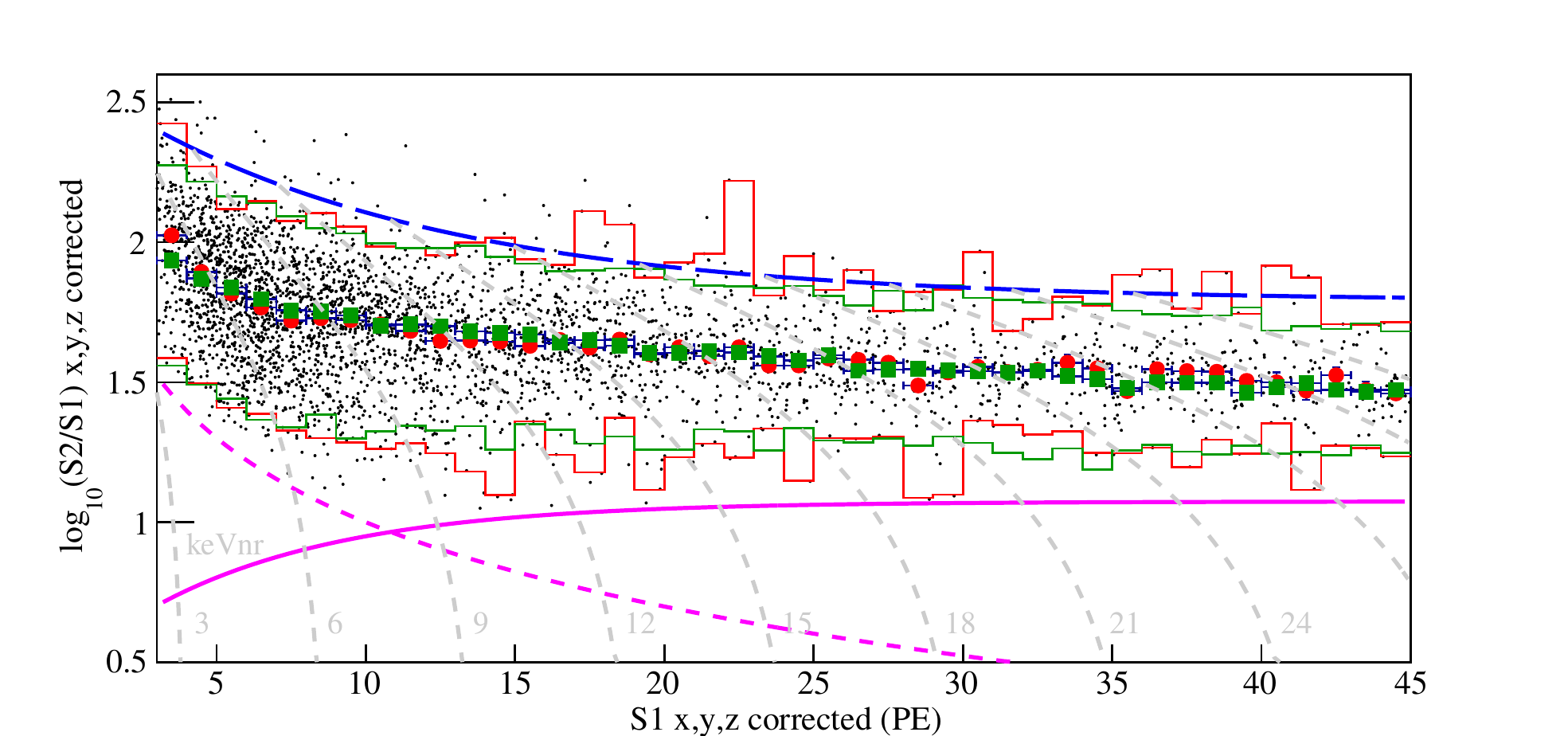}
  \end{subfigure}
  \caption{Top: tritium calibration
    data in $\log_{10}(\text{S2/S1})$ vs. S1, and fits of medians of ER (blue)
    and NR (red) data.
    Bottom: AmBe calibration data in $\log_{10}(\text{S2/S1})$ vs. S1, together with 
    medians from the data (red solid circles) and MC (green squares), and the 
    fit to ER medians (blue dashed). The 
    2.3- and 97.7-percentiles from the data (red lines) and MC (green lines) 
    are overlaid. The dashed magenta curve is the 100 PE selection cut 
    for S2. In both panels, solid magenta curves represent the 
    99.9\% NR acceptance curve from the MC. 
    The gray dashed curves represent the equal energy curves in ER and NR energy 
    for the top and bottom panels, respectively.
}
\label{fig:calibration}
\end{figure}

\begin{figure}[h]
  \centering
  \includegraphics[width=0.45\textwidth]{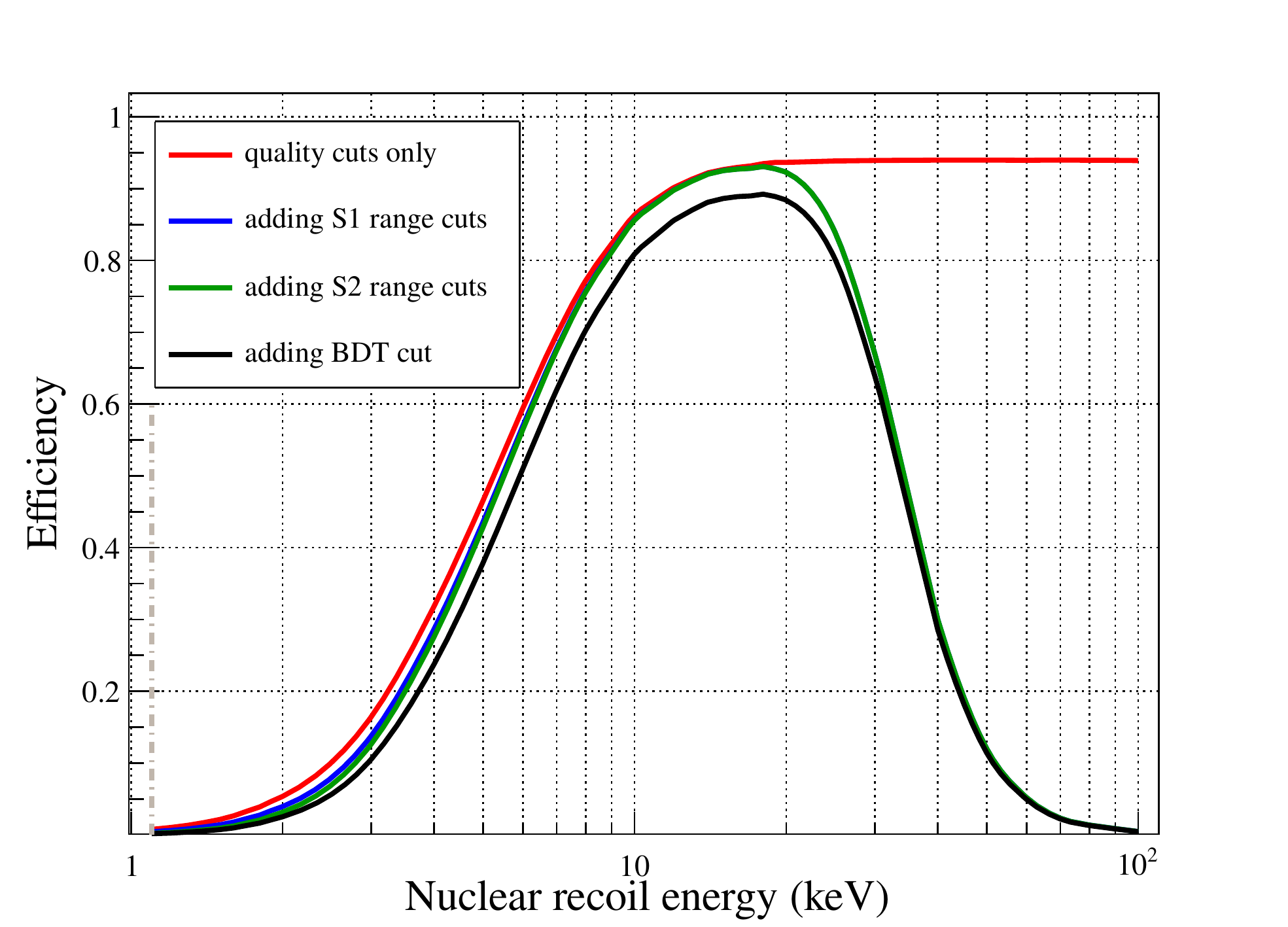}
  \caption{
    The detection efficiencies as functions of the NR energy using 
    untuned NEST model after successive applications of selections 
    indicated in the legend. The BDT acceptance shown is determined 
    by applying the BDT efficiency as a function of S1; the acceptance 
    is virtually unchanged if the BDT efficiency is instead applied to S2.
    The dashed line at 1.1 keV$_{nr}$ indicates 
    the cutoff used in the WIMP limit setting.
  }
  \label{fig:eff_curves}
\end{figure}

The ER and NR calibration events through tritiated methane and AmBe sources are 
shown in $\log_{10}(\text{S2/S1})$ vs.  S1 in Fig.~\ref{fig:calibration} after 
the dark matter selection cuts (Table~\ref{tab:eventrate}).
The AmBe calibration was carried out in-between the 
dark matter running periods.
A full Monte Carlo (MC) simulation based on 
Geant4~\cite{Agostinelli:2002hh, Allison:2006ve} (v10.2) including
neutron-gamma correlated emission from the source, detailed detector
geometry, neutron propagation in and interactions with the
detector (following the recommended physics list~\cite{g4_physics_reference}), 
and S1-S2 signal production (NEST-1.0-based~\cite{Lenardo:2014cva}) 
was developed to compare to the
data. According to the simulation, less than 10\% of the low energy events are
contaminated with multiple scattering in the dead region
(``neutron-X''). The data quality cuts mentioned above further suppress
the neutron-X events, therefore the final sample was
considered as a pure single-scatter NR sample at this stage.
3447 such events were collected for 6.8 days in total, 
with less than 1\% contamination 
from the ER background.
In the bottom panel of 
Fig.~\ref{fig:calibration}, the medians and widths of the
NR data were compared to those obtained from the
NEST-1.0 NR model~\cite{Lenardo:2014cva} with detector parameters
(PDE, EEE, SEG, and DPF) taken into account, and a good
agreement is observed. The distributions in S1 and S2 
were also compared to the MC, a small discrepancy was observed in detailed shape 
for small S1 and S2. A tuning of the NEST model 
could improve the comparison (see Fig.~\ref{fig:nr_comb} in Supplemental Material [\cite{sup_material}]) but worsen the 
agreement for medians of log$_{10}$(S2/S1) at low energy, which 
warrant further investigation. Following the standard 
practice, untuned NEST is used when reporting the official results.
The corresponding efficiencies as
functions of NR energy are shown in Fig.~\ref{fig:eff_curves}, which are
later applied to calculate the dark matter detection
efficiency.

At the end of Run 9, tritiated methane with a specific activity of 0.1 mCi/mmol was
administered into the detector through a liquid-nitrogen cold trap, a
leak valve, and a 100 mL sample chamber under vacuum, which was
flushed with xenon gas in the detector. We collected a total of
2807 tritium $\beta$-decay events. Among these, 9 leaked below the
median of the NR band, leading to a leakage fraction of
(0.32$\pm$0.11)\%.

During the krypton distillation campaign in early 2016, 1.1-ton of
xenon was exposed to about one month of sea level cosmic ray
radiation, leading to the production of $^{127}$Xe, which then decayed
via electron capture (EC) to $^{127}$I producing characteristic ER
energy deposition in the detector.  The $^{127}$Xe level was
identified by the 33 keV K-shell X-ray (following EC), 
with a decay rate of about 1.1$\pm$0.3 and 0.1$\pm$0.03 mBq/kg at the beginning
and end of Run 9, respectively. In the
low-energy region, M-shell and L-shell vacancies of $^{127}$I can
produce 1.1 keV and 5.2 keV ER events in the detector, repectively.
The background was estimated to be 0.37$\pm$0.05 mDRU (1 mDRU =
$10^{-3}$ events/day/kg/keV$_{ee}$ where keV$_{ee}$ represents
``electron equivalent'' energy) below 10 keV$_{ee}$ based on the MC by
scaling the measured K-shell X-ray rate, in good agreement with
0.42$\pm$0.08 and 0.40$\pm$0.13 mDRU obtained from the spectrum fit
and time-dependence fit of the low energy events. 
We chose 0.42 mDRU as 
the nominal value with a systematic uncertainty of 25\%. The krypton
background level was estimated {\it in-situ} using the
$\beta$-$\gamma$ delayed coincidence from $^{85}$Kr decay. In total,
52 candidates were identified in Run 9 within 329 kg of FV (discussed later) 
and no time dependence was
observed, leading to an estimate of 44.5$\pm$6.2 ppt of Kr in xenon
assuming a $^{85}$Kr concentration of 2$\times10^{-11}$ in natural
Kr. This represents a factor of ten reduction compared to the Kr level
in Ref.~\cite{Tan:2016diz}.

The backgrounds due to radio-impurity of detector components shall be
the same as those in Ref.~\cite{Tan:2016diz}, and so is the neutron
background. The ER background due to Rn was estimated {\it in-situ}
using the $\beta$-$\alpha$ and $\alpha$-$\alpha$ delayed coincidence
events.  The $^{222}$Rn and $^{220}$Rn decays were estimated to be
8.6$\pm$4.6 and 0.38$\pm$0.21 $\mu$Bq/kg, respectively, consistent
with the results in Ref.~\cite{Tan:2016diz}. We also estimated the
background from the $^{136}$Xe double-$\beta$ decay events and
neutrinos (see Ref.~\cite{Billard:2013qya} and references therein).
The former produces an ER background of 0.10$\pm$0.01 events per 10000
kg-day. The neutrino ER background is dominated by pp solar
neutrinos, and is estimated to be between 0.2 and 6.0 events per 10000
kg-day where the lower and upper values assume zero or the current
experimental limit of the neutrino magnetic
moment~\cite{Beda:2013mta}, respectively. The neutrino NR background
was estimated to be 1$\times10^{-3}$ events per 10000 kg-day. The final
low energy background composition is summarized in
Table~\ref{tab:er_bkg_budget}.

\begin{table}[h]
  \begin{tabular}{ccc}
    \hline\hline
    Item & Run 8 (mDRU) & Run 9 (mDRU)\\\hline
    $^{85}$Kr & 11.7 & 1.19 \\
    $^{127}$Xe & 0 & 0.42 \\
    $^{222}$Rn & 0.06 & 0.13  \\
    $^{220}$Rn & 0.02 & 0.01  \\
    Detector material ER & 0.20 & 0.20\\
    \hline
    Total & 12.0 & 1.95\\\hline
  \end{tabular}
  \caption{Summary of ER backgrounds from different components in
    Run 8 and Run 9. The fractional uncertainties for $^{85}$Kr and $^{127}$Xe
    are 17\% and 0\% for Run 8 and 14\% and 25\% for Run 9, respectively.
    The uncertainties for $^{222}$Rn and $^{220}$Rn in both runs are
    taken to be 54\% and 55\%, respectively. The fractional
    uncertainty due to detector materials is estimated to be 50\% based on the
    systematic uncertainty of the absolute efficiency of the gamma
    counting station. Different from Ref.~\cite{Tan:2016diz}, values in
    the table are now folded with detection efficiency.
  }
  \label{tab:er_bkg_budget}
\end{table}

Similar to Ref.~\cite{Xiao:2015psa}, 
the accidental background was computed by randomly
pairing the isolated S1 (1.8 Hz) and S2 (approximately
1500/day) events within the dark matter selection range.
The data quality cuts mentioned earlier suppressed this 
background to 33\%, among which 15\% is below the NR median.
To optimize the rejection for such background, 
further cuts were developed based on the boosted-decision-tree
(BDT) method~\cite{Roe:2004na}, in which 
the below-NR-median AmBe calibration data and randomly paired
S1-S2 signals were used as the input signal and
background, respectively. 
The input data were split into two equal
statistics sets, one for training and the other for test. 
The BDT cuts target, for example, events where the drift time and S2-width 
are inconsistent, or where the S2 shape indicates an S2 originating from the 
gate or gas regions.
13 variables entered into the training including the charge of S1, S2 and the drift time,
the width, 10\%-width, rising slope, waveform asymmetry, and top/bottom ratio of the S2, 
ratio of maximum bottom channel to total and top/bottom ratio of the S1, 
spikes-within and spikes-around the S1 (indicators of a S2 mis-ID), 
and the pre-S2 area in a window between 1.5 and 3 $\mu$s 
(tag of a gate event).
After applying the BDT cuts to the accidental background surviving the data quality cuts,
events below the NR median were strongly 
suppressed to 27\%, while the overall AmBe NR efficiency
was maintained at 93\%.
The uncertainty on the remaining accidental background was estimated to
be 45\% using the difference found in Run 8 and Run 9.

Similar to Ref.~\cite{Tan:2016diz}, the final S1 range cut was chosen
to be between 3 to 45 PE, corresponding to an average energy window
between 1.3 to 8.7 keV$_{ee}$ (4.6 to 35.0 keV$_{nr}$), 
and S2s were required to be between 100 PE (raw) and 10000 PE (uniformity corrected). 
For all events with a
single S2, the FV cut was determined based on the PAF-reconstructed
position distribution. 
The selection criterion in the horizontal plane was
taken to be $r<$268~mm, using
data with S1 outside of the dark matter search window (between 50 and 200 PE).
The drift time was required to be
between 18 to 310 $\mu$s, 
where the maximum drift time cut was to suppress the
below-cathode $\gamma$ energy deposition (so-called ``gamma-X'') from
$^{127}$Xe decays.  
The liquid xenon mass was estimated to be
329$\pm$16 kg, where the uncertainty was estimated based on the
position difference between the PAF and TM methods, consistent 
with other estimates using the tritium event distribution 
and expected intrinsic resolution from the TM method.
The vertical electric field deformation resulting from the accumulations of wall 
charges in the TPC~\cite{Akerib:2015rjg} was estimated using the $^{210}$Po 
plate-out events from the PTFE wall. 
The reconstructed positions are 2.4 mm (18~$\mu$s) to 6.3 mm (310~$\mu$s) away 
from the geometrical wall, a combined effect from reconstruction and field distortion. 
Therefore we have made a conservative choice of the FV and 
neglected the field deformation therein.
Under these cuts, the final expected background budget is
summarized in Table~\ref{tab:backgroundtable}.

\begin{table}[h]
\centering
\begin{tabular}{cccc|c|c}
\hline\hline
 & ER & Accidental & Neutron & \parbox[t]{1.4cm}{Total\\Expected} & \parbox[t]{1.4cm}{Total\\observed}\\
\hline
Run 8 & 622.8 & 5.20 & 0.25 & 628$\pm$106 & 734 \\
\hline
\parbox{1.8cm}{Below \\NR median} & 2.0 & 0.33 & 0.09 & 2.4$\pm$0.8 & 2\\
\hline
Run 9 & 377.9 & 14.0 & 0.91 & 393$\pm$46 & 389 \\
\hline
\parbox{1.8cm}{Below \\NR median} & 1.2 & 0.84 & 0.35 & 2.4$\pm$0.7 & 1\\
\hline
\hline
\end{tabular}
\caption{The expected background events in Run 8 and Run 9
  in the FV, before and after the NR median cut.
  The fractional uncertainties of expected events in the table
  are 17\% (Run 8 ER), 12\% (Run 9 ER),  45\% (accidental), and 
  100\% (neutron), respectively. 
  Both the uncertainties from the ER rate and leakage
  fraction, (0.32$\pm$0.11)\%, were taken into account
  in estimating the uncertainty of ER background below the NR median.
 Number of events from the data are shown in the
 last column.}
\label{tab:backgroundtable}
\end{table}

\begin{figure}
  \centering
  \includegraphics[width=0.45\textwidth]{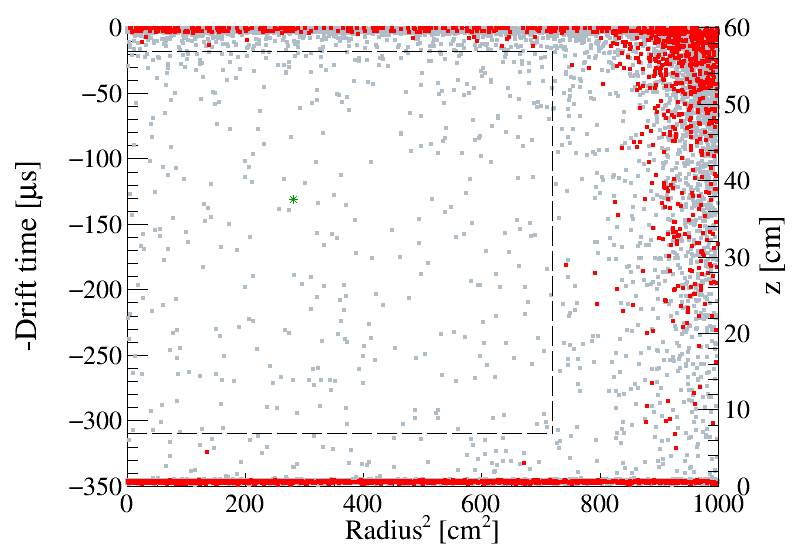}
  \caption{Position distribution of events that pass all selections (gray points),
    and those below the NR median (outside FV: red points; inside FV: green star),
    with FV cuts indicated as the black dashed box.}
  \label{fig:events_pos}
\end{figure}

\begin{figure}[h]
  \centering
  \includegraphics[width=0.45\textwidth]{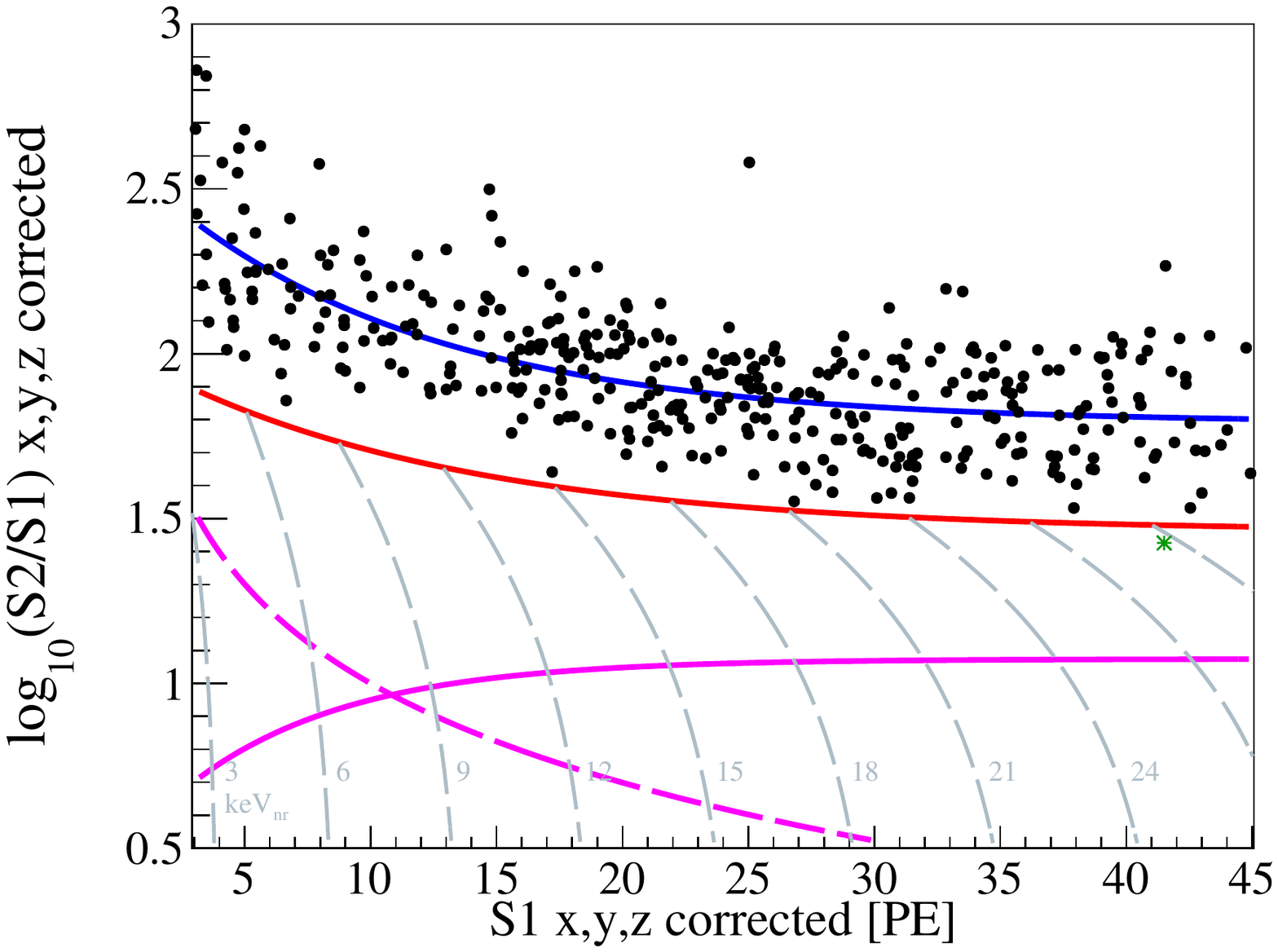}
  \caption{The distribution of $\log_{10}$(S2/S1) versus S1 for the dark matter search data.
    The median of the NR calibration
    band is indicated as the red curve. The dashed magenta curve represents the equivalent
    100~PE cut on S2. The solid magenta curve is the 99.99\% NR acceptance curve.
    The gray dashed curves represent the equal energy curves
    with NR energy indicated in the figures. The data point below the NR median
    curve is highlighted as a green star.}
  \label{fig:dm_band}
\end{figure}

The event rates of Run 9 after successive selections are summarized
in Table~\ref{tab:eventrate}. The skin veto selections are more effective
than that in Run 8 since the background was less dominated by the
volume-uniform $^{85}$Kr $\beta$-decays. 
The vertex distribution of all events before and after the
FV cut is shown in Fig.~\ref{fig:events_pos}. 
Outside the FV, pile-up of events near the cathode, the gate and the wall were observed.
After the FV cut, 389 events
survived, and event distribution in 
radius-square agree statistically with a flat distribution, 
indicating no effects from the electric field deformation due to wall charges.
One event was found below the
NR median curve, with its location indicated in
Fig.~\ref{fig:events_pos}. 
The $\log_{10}$(S2/S1) vs. S1
distribution for the 389 candidates is shown in
Fig.~\ref{fig:dm_band}. Being close to the NR median line, the single
below-NR-median event is consistent with a leaked ER background.

\begin{table}[h]
\centering
\begin{tabular}{ccc}
\hline\hline
Cut & \#Events & Rate (Hz) \\
\hline
All triggers & 24502402 & 3.56 \\
Single S2 cut & 9783090 & 1.42 \\
Quality cut & 5853125 & 0.85 \\
Skin veto cut & 5160513 & 0.75 \\
S1 range  & 197208 & 2.87$\times 10^{-2}$ \\
S2 range  & 131097 & 1.91$\times 10^{-2}$ \\\hline
18 $\mu$s FV cut & 21079 & 3.06$\times 10^{-3}$\\
310 $\mu$s FV cut  & 7361 & 1.07$\times 10^{-3}$\\
268 mm FV cut & 398  & 5.79$\times 10^{-5}$ \\\hline
BDT cut & 389 & 5.66$\times 10^{-5}$\\
\hline\hline
\end{tabular}
\caption{The event rates in Run 9 after various analysis selections.}
\label{tab:eventrate}
\end{table}

The data in Run 8 and Run 9 were combined in the final analysis to obtain
a new WIMP search limit. The Run 8 data were reanalyzed with the
updated reconstruction and data selection cuts except that the
vertical cuts were maintained from 20 to 346 $\mu$s (FV = 367 kg),
since there was no gamma-X contamination from $^{127}$Xe in Run 8.
This represents the largest dark-matter-search data set among
dual-phase xenon detectors to date with an overall exposure of
3.3$\times10^{4}$ kg-day.  A likelihood approach similar to that in
Ref.~\cite{Xiao:2015psa} was used to fit the measured data
distribution in S1 and S2. A parameterized tritium event distribution 
was used to simulate expected distributions for different ER background
components, and that for accidentals was obtained from the data. The DM 
NR signals were simulated with the untuned NEST for different WIMP 
masses as the expected DM distributions, with the conservative low energy 
cutoff at 1.1~keV$_{nr}$~\cite{Akerib:2015rjg}. 
The entire data set was separated into 15 time bins to 
take into account the time dependent factors such as the electron lifetime, 
the background level, and other detector parameters (Table~\ref{tab:run_sets}).
The scales of all five
background components, $^{127}$Xe, $^{85}$Kr, other ER background (including 
Rn and material background), accidental, and
neutron background, were defined as global nuisance parameters with their corresponding
Gaussian penalty terms constructed based on systematic uncertainties 
in Tables~\ref{tab:er_bkg_budget} and \ref{tab:backgroundtable}. The nominal 
rates for the latter four backgrounds were taken from Table~\ref{tab:er_bkg_budget}, and for 
$^{127}$Xe, the nominal rate was derived from the table to include the time dependence. 
The uncertainties for PDE, EEE, and SEG in the caption of Table~\ref{tab:run_sets}
were verified to have small impact 
to the results and were neglected in the likelihood fit for simplicity.

\begin{figure}[h]
  \centering
  \includegraphics[width=0.45\textwidth]{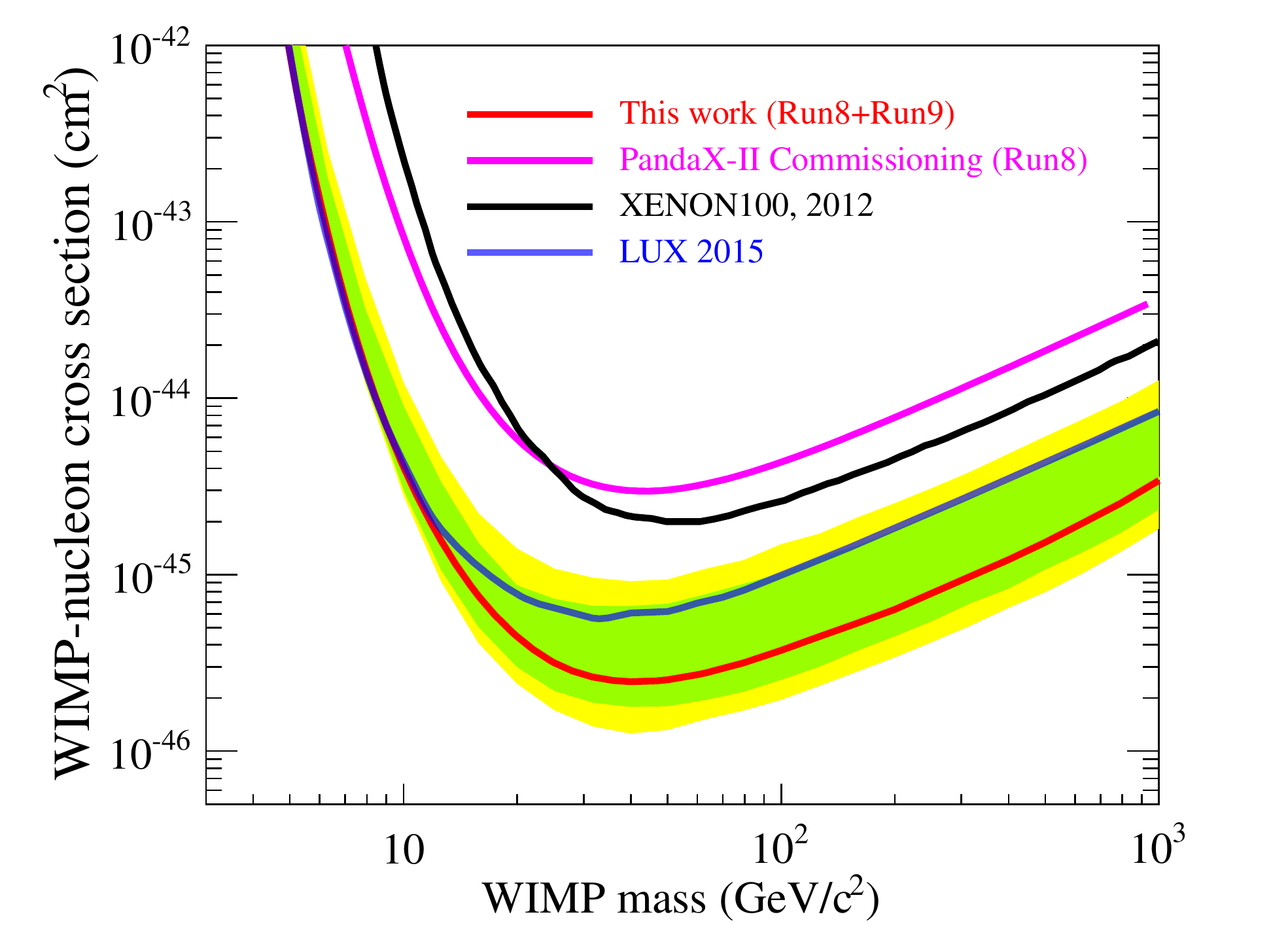}
  \caption{The 90\% C.L. upper limits for the spin-independent isoscalar
WIMP-nucleon cross sections from the combination of PandaX-II Run 8
and Run 9 (red solid).
Selected recent world results are plotted for comparison:
PandaX-II Run 8 results~\cite{Tan:2016diz} (magenta),
XENON100 225 day results~\cite{Aprile:2013doa} (black), and
LUX 2015 results~\cite{Akerib:2015rjg}(blue). The 1 and 2-$\sigma$
sensitivity bands are shown in green and yellow, respectively.}
\label{fig:limit}
\end{figure}

To obtain the exclusion limit to spin-independent isoscalar
WIMP-nucleon cross section, profile likelihood ratio~\cite{Cowan:2010js,Aprile:2011hx} was constructed over
grids of WIMP mass and cross section, and the final 90\% confidence
level (C.L.) cross section upper limits were calculated using the
CL$_{\mathrm{s}}$ approach~\cite{CLS1,CLS2}. The final results are
shown in Fig.~\ref{fig:limit}, with recent results from PandaX-II Run
8~\cite{Tan:2016diz} and LUX~\cite{Akerib:2015rjg} overlaid. 
Our upper limits lie within the
$\pm$1$\sigma$ sensitivity band, consistent with statistical expectation based on 
Table~\ref{tab:backgroundtable}. The lowest cross section limit
obtained is 2.5$\times$10$^{-46}$~cm$^2$ at a WIMP mass of 40
GeV/c$^2$, which represents an improvement of more than a factor of 10
from Ref~\cite{Tan:2016diz}. In the high WIMP mass region, our results
are more than a factor of 2 more stringent than the LUX
results~\cite{Akerib:2015rjg}. Note that we have been generally
conservative in officially reporting the first limits in this article.
WIMP NR modeling with a tuned NEST could result in an even more 
stringent limit (see
Fig.~\ref{fig:limits_tuned} in Supplemental Material [\cite{sup_material}]), and a 
more elaborated treatment of FV cuts would also help. 

In conclusion, we report the combined WIMP search results using data
from Run 8 and Run 9 of the PandaX-II experiment with an exposure of
3.3$\times$10$^4$~kg-day. No dark matter candidates were identified
above background and 90\% upper limits were set on the
spin-independent elastic WIMP-nucleon cross sections with a lowest
excluded value of 2.5$\times$10$^{-46}$~cm$^2$ at a WIMP mass of
40~GeV/c$^2$, the world best reported limit so far. 
The result is complementary to the searches performed at the LHC, which have produced 
various WIMP-nucleon cross section limit in the range from 10$^{-40}$
to 10$^{-50}$ (c.f. Refs.~\cite{Aad:2015pla} and \cite{Aad:2015zva}),
dependent on the dark matter production models.
The PandaX-II experiment
continues to take physics data to explore the previously unattainable
WIMP parameter space.

\begin{acknowledgments}
  This project has been supported by a 985-III grant from Shanghai
  Jiao Tong University, grants from National Science Foundation of
  China (Nos. 11435008, 11455001, 11505112 and 11525522), and a grant
  from the Office of Science and Technology in Shanghai Municipal
  Government (No. 11DZ2260700). This work is supported in part by the
  Chinese Academy of Sciences Center for Excellence in Particle
  Physics (CCEPP).  The project is also sponsored by Shandong
  University, Peking University, and the University of Maryland.  We
  also would like to thank Dr. Xunhua Yuan and Chunfa Yao of China
  Iron \& Steel Research Institute Group, and we are particularly
  indebted to Director De Yin from Taiyuan Iron \& Steel (Group)
  Co. LTD for crucial help on nuclear-grade steel plates. We also
  thank C. Hall for helping with specifics of tritium
  calibration. Finally, we thank the following organizations and
  personnel for indispensable logistics and other supports: the CJPL
  administration including directors Jianping Cheng and Kejun Kang and
  manager Jianmin Li, and the Yalong River Hydropower Development
  Company Ltd.
\end{acknowledgments}

\bibliographystyle{apsrev4-1}
\bibliography{refs.bib}

\onecolumngrid
\appendix
\section{Supplementary Materials} 
\begin{figure}[H]
  \centering
  \includegraphics[width=.45\textwidth]{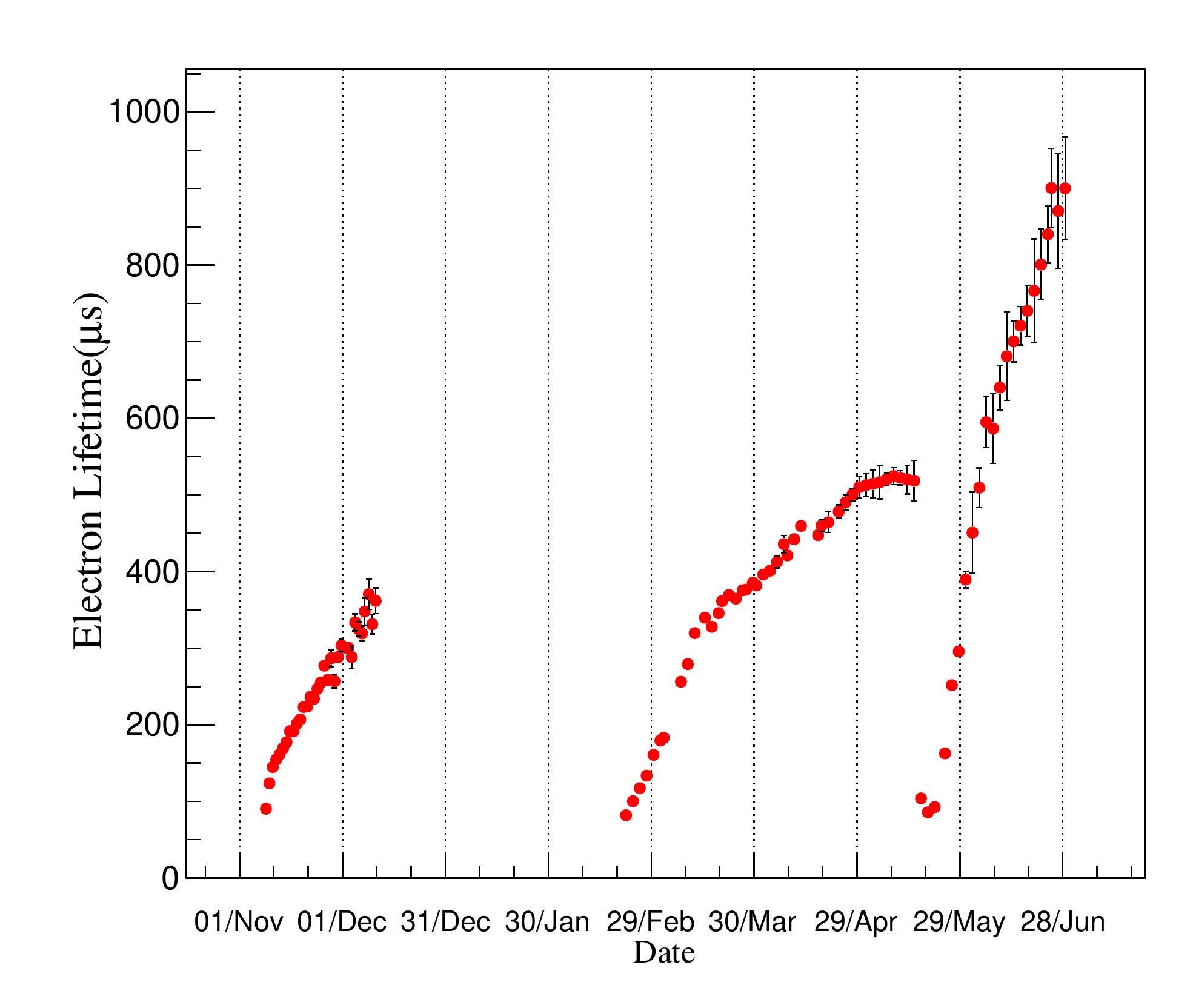}
  \caption{Evolution of the electron lifetime in Run 8 and Run 9. Each point represent
    the average in a data taking unit, usually lasted for 1 or 2 days. Only data 
    with electron lifetime longer than 205 $\mu$s were used in 
    the dark matter analysis.}
  \label{fig:e_lifetime}
\end{figure}

\begin{figure}[H]
  \centering
  \includegraphics[width=.45\textwidth]{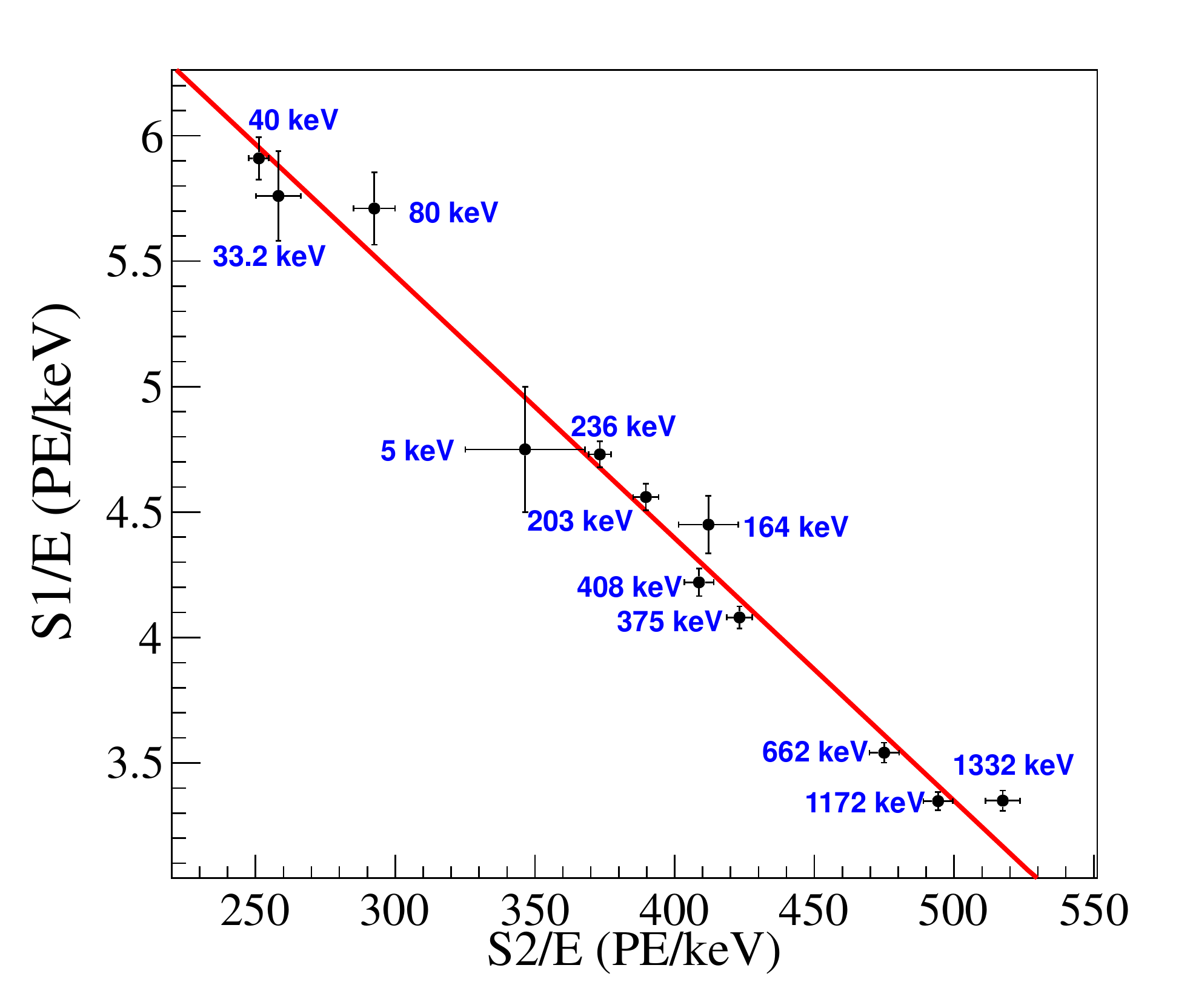}
  \caption{Linear fit in S2$/E$ vs. S1$/E$ for all ER peaks in data to determine the 
    PDE and EEE. S2 and S1 were obtained from Gaussian fits to each S2 and S1 peak, respectively, and only statistical uncertainties from the fits are shown.}
  \label{fig:doke_plot}
\end{figure}

\begin{figure}[H]
  \centering
  \includegraphics[width=.45\textwidth]{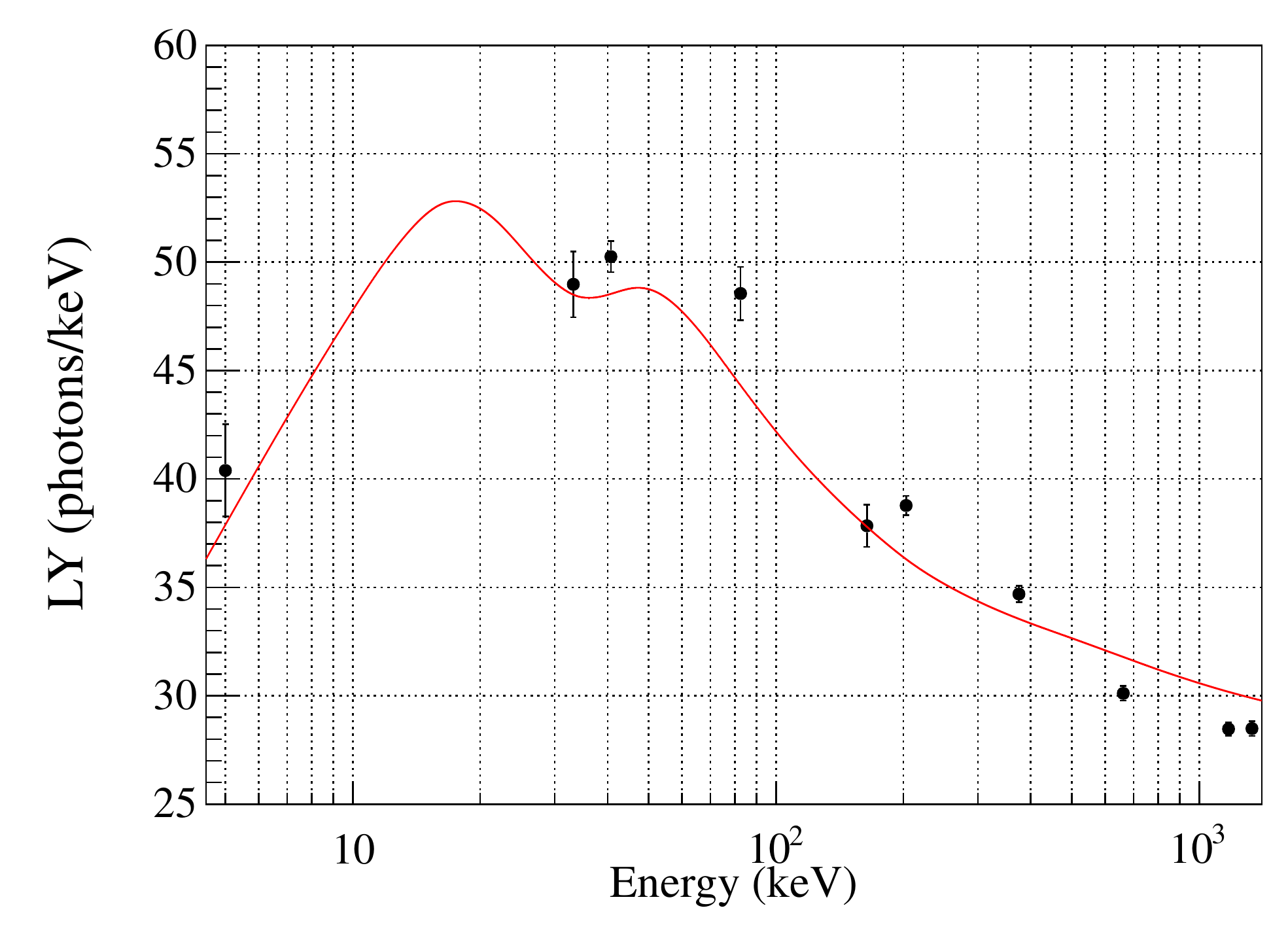}
  \includegraphics[width=.45\textwidth]{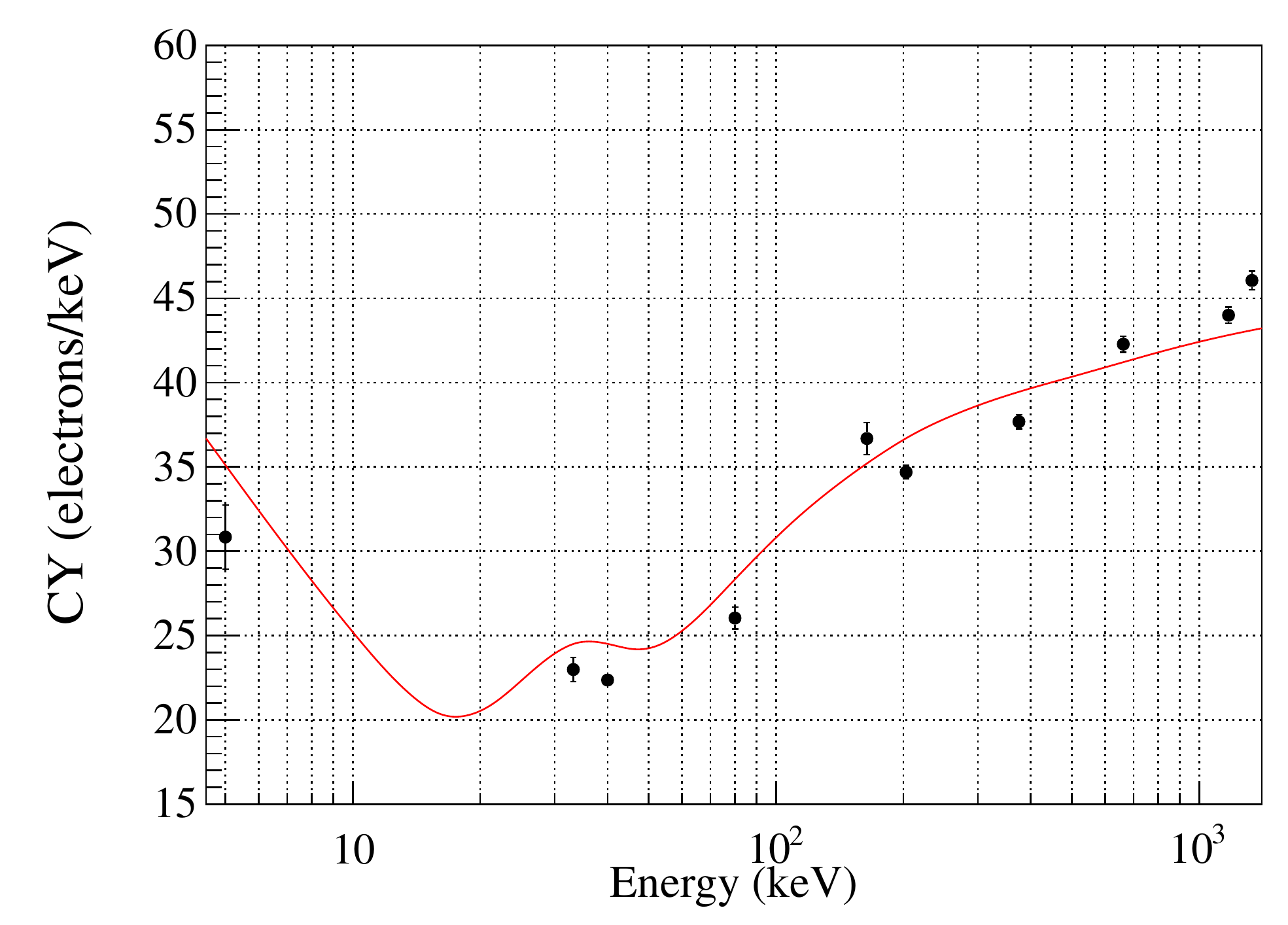}
  \caption{Comparison of measured ER light yield (left) and charge yield (right) with NEST predictions. Only statistical uncertainties are shown. The systematic uncertainties of PDE
 and EEE are estimated by the difference between the data and NEST predictions.}
  \label{fig:nest_comparison}
\end{figure}

\begin{figure}[H]
  \centering
  \includegraphics[width=.45\textwidth]{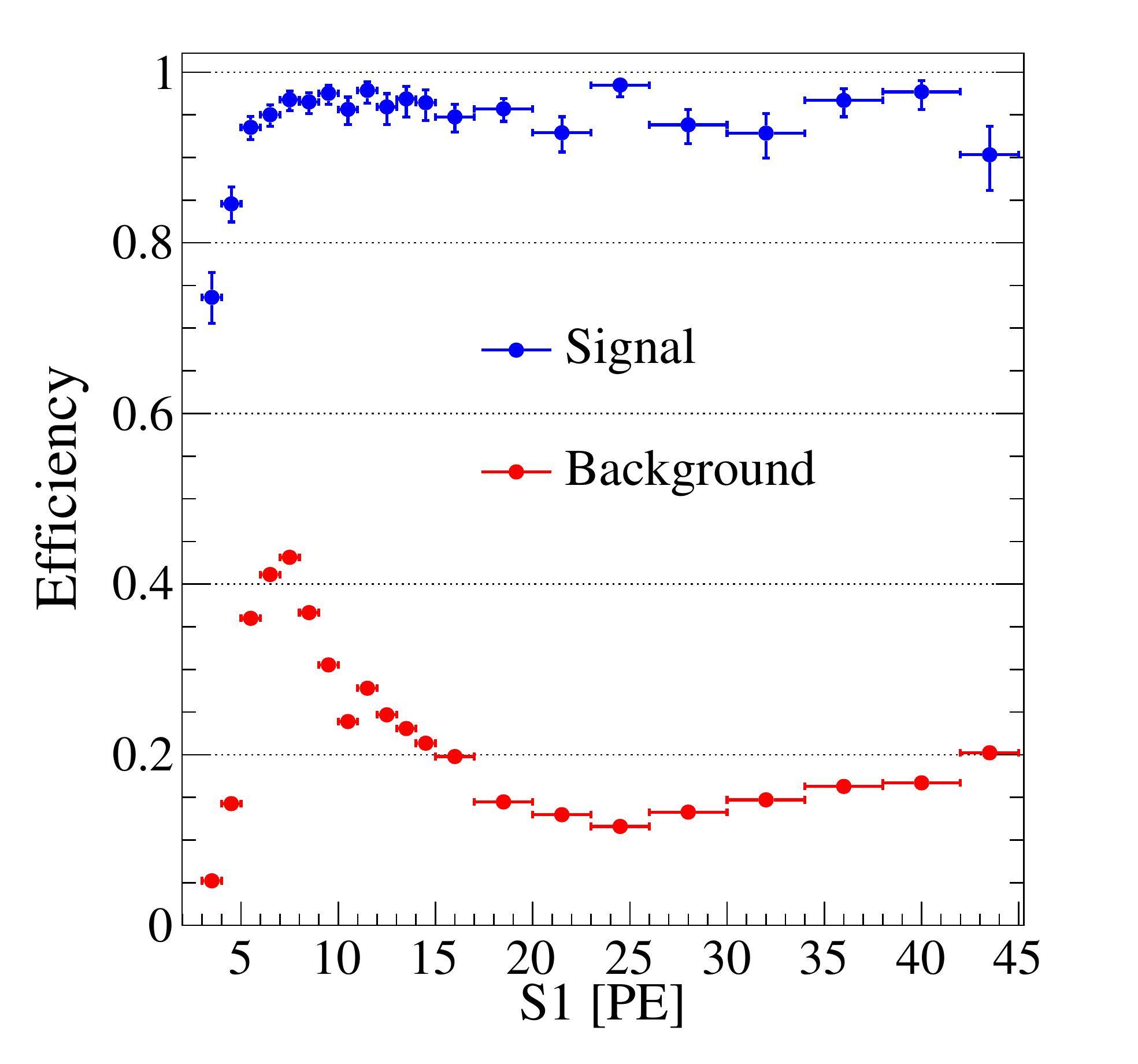}
  \includegraphics[width=.45\textwidth]{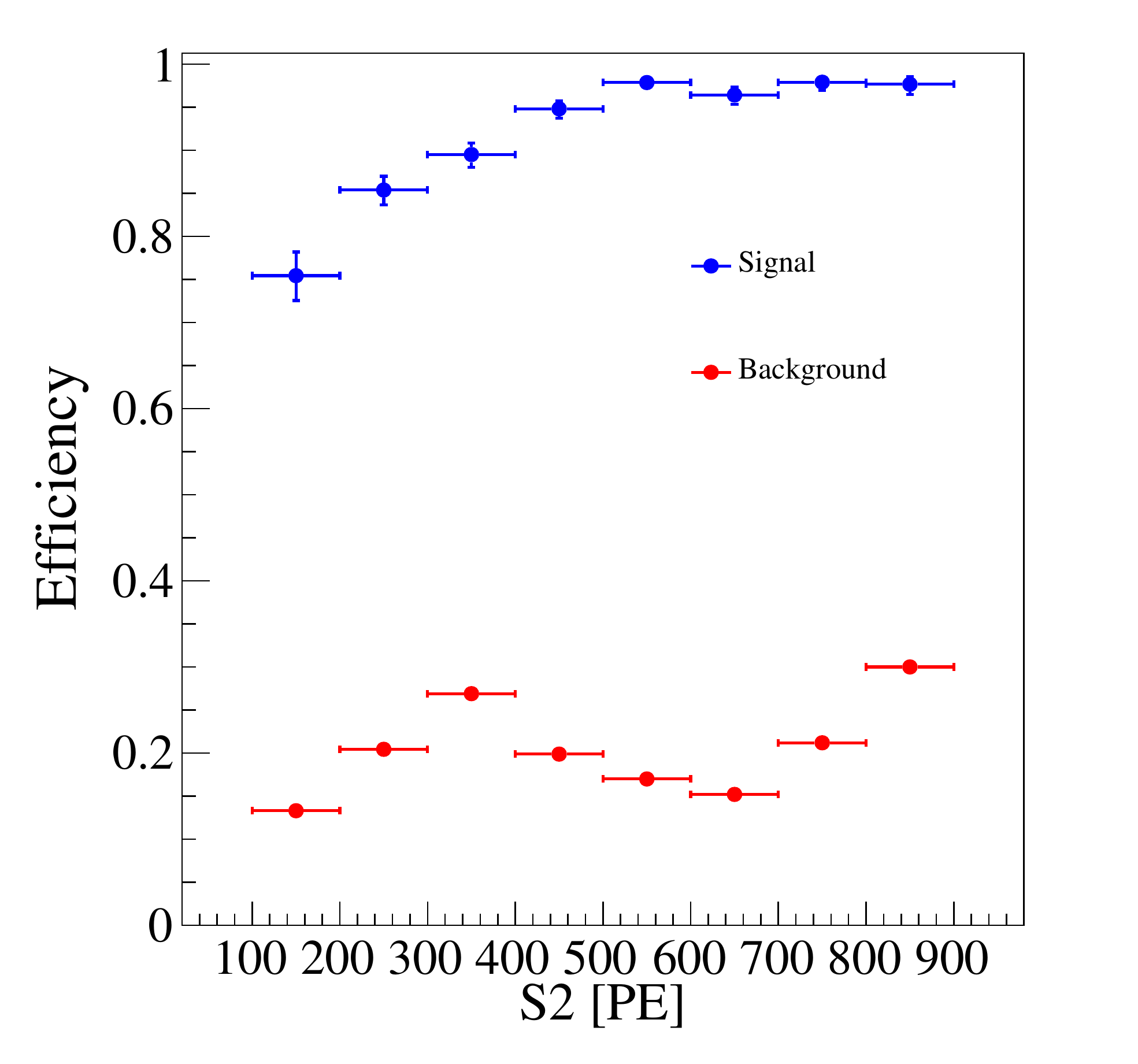}
  \caption{The BDT efficiency for the signal (NR) and background (accidental) events, both
    are selected below the NR median, projected to S1 (left) and S2 (right).}
  \label{fig:bdt_vs_s1}
\end{figure}

\begin{figure}[H]
  \centering
  \begin{subfigure}[b]{0.45\textwidth}
    \includegraphics[width=\textwidth]{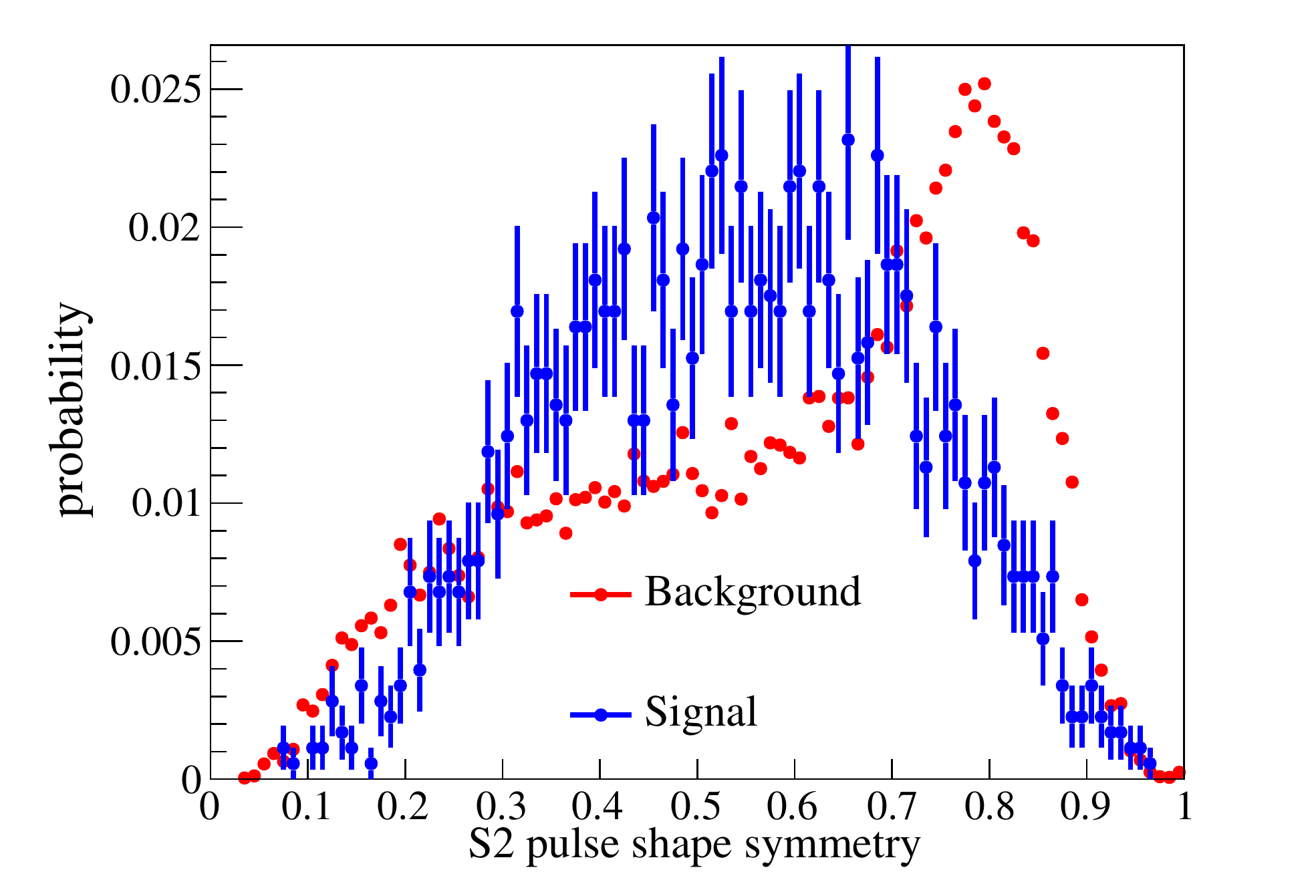} 
  \end{subfigure}
  \begin{subfigure}[b]{0.45\textwidth}
    \includegraphics[width=\textwidth]{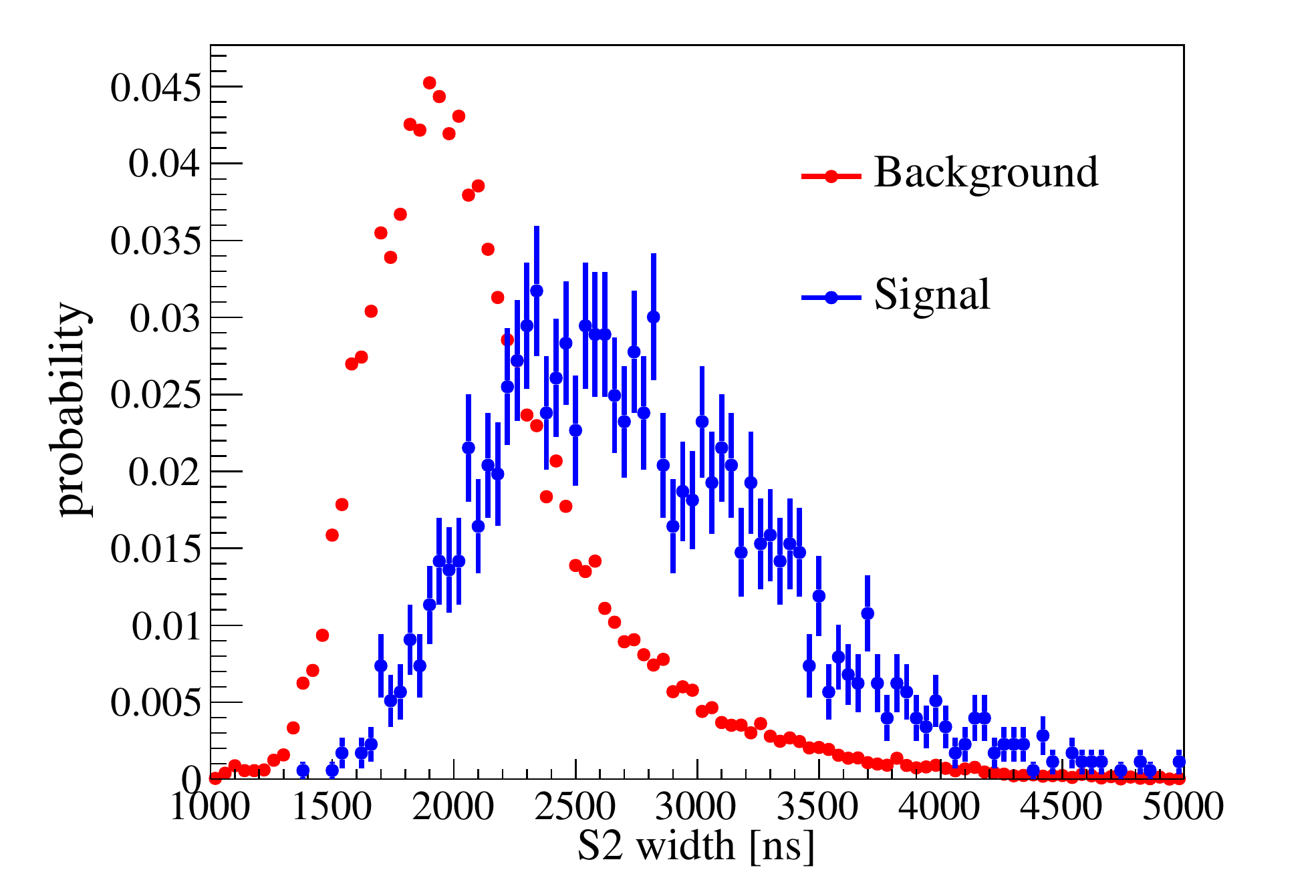} 
  \end{subfigure}
  \caption{Examples of distributions of the input variables used for BDT 
    in the signal and background training
    samples. Left: S2 pulse shape symmetry, defined as 
    ratio of pre-peak area to the total area of an S2, Right: the width of S2.}
  \label{fig:bdt_parameters}
\end{figure}

\begin{figure}[H]
  \centering
  \includegraphics[width=.3\textwidth]{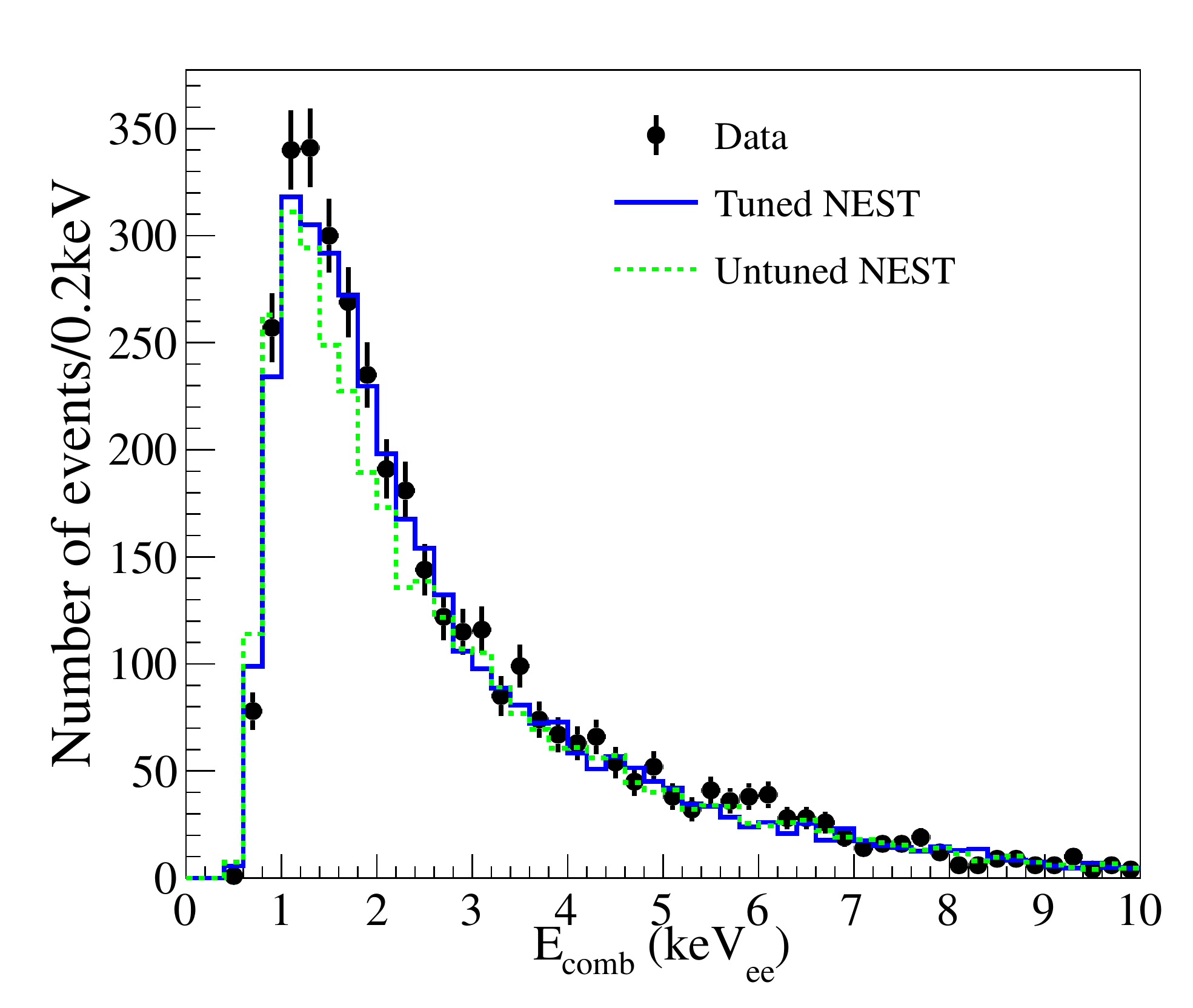}
  \includegraphics[width=.3\textwidth]{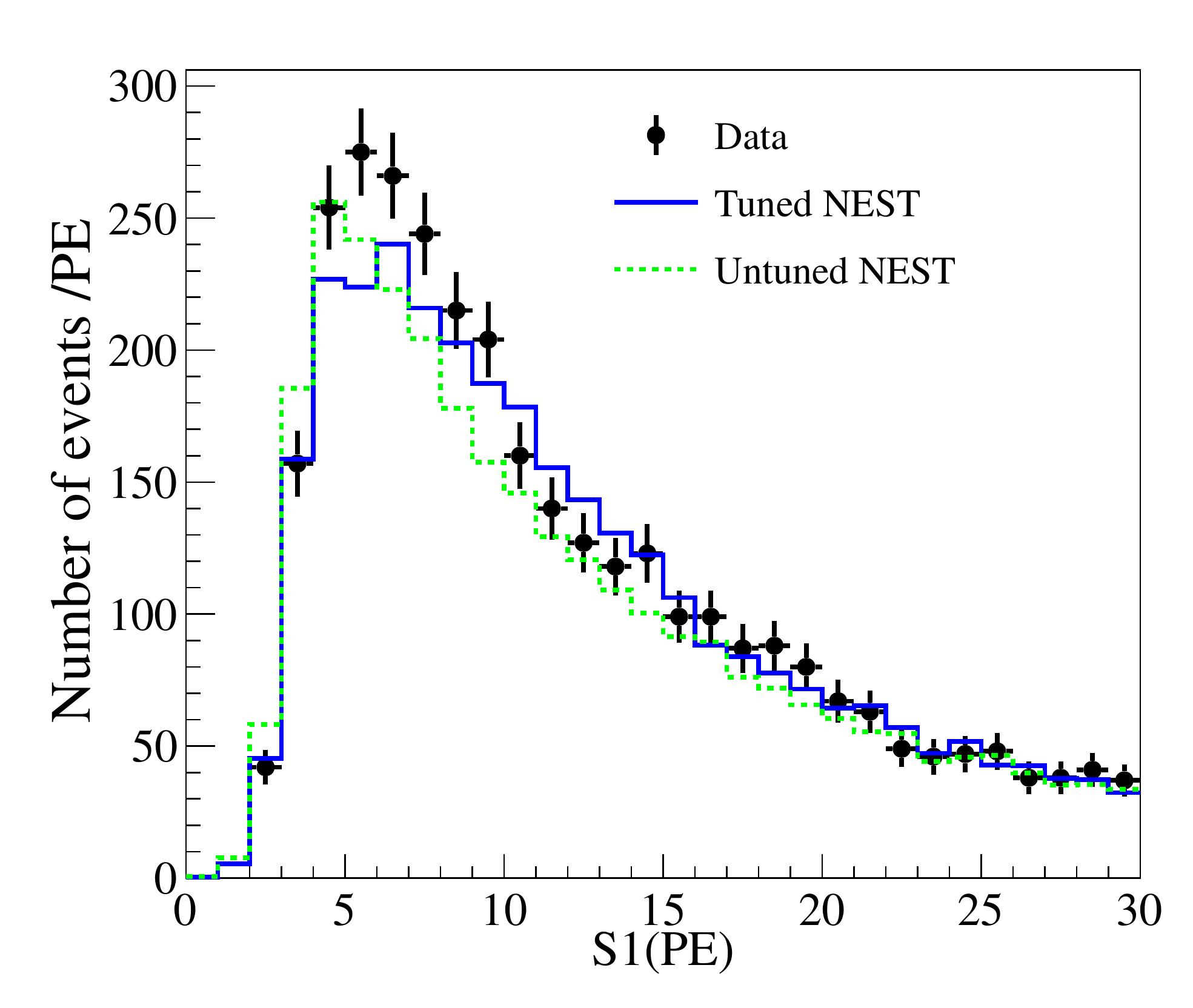}
  \includegraphics[width=.3\textwidth]{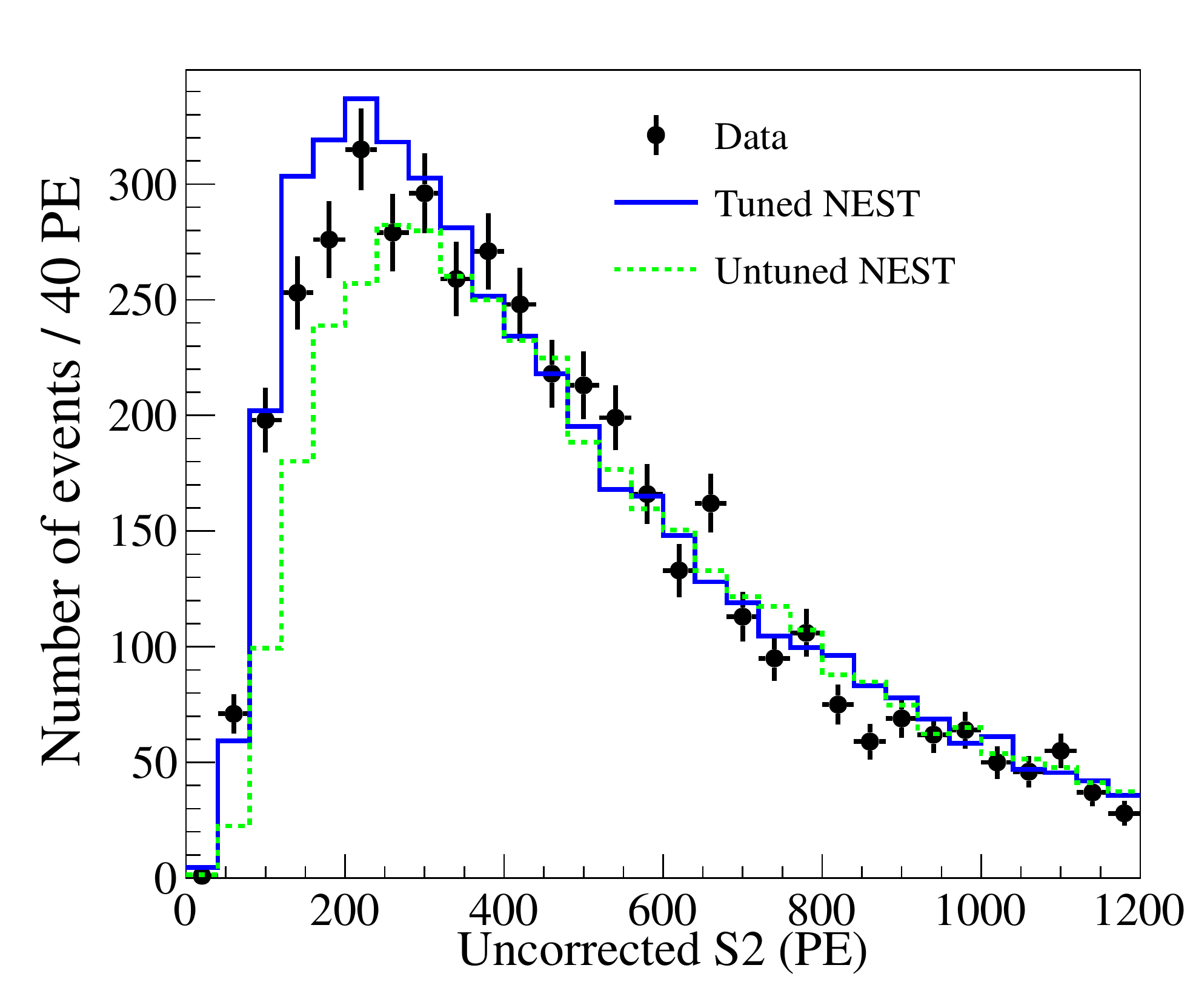}
  \caption{Comparison of distributions between the AmBe data and MC (untuned and 
    tuned, with detection efficiency applied) in combined energy in keV$_{ee}$ (left), 
    S1 (middle) and raw S2 (right).
  }
  \label{fig:nr_comb}
\end{figure}

\begin{figure}
\centering
\includegraphics[width=0.45\textwidth]{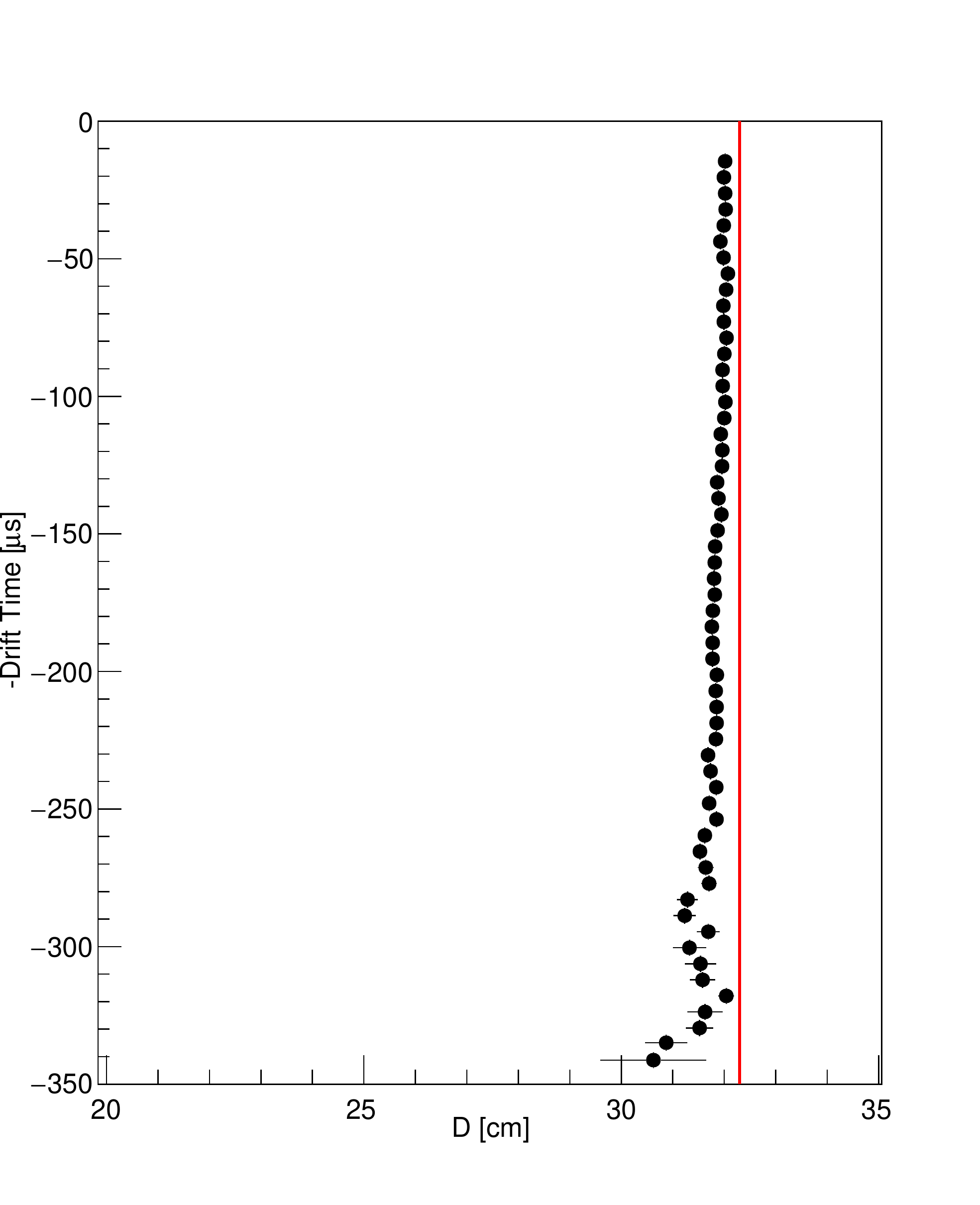}
\caption{Reconstructed radius vs. drift time for the $^{210}$Po plate-out events from the 
  PTFE wall. The location of the PTFE wall is indicated as the red line.}
\label{fig:wall}
\end{figure}

\begin{figure}[H]
  \centering
  \begin{subfigure}[b]{0.45\textwidth}
  \includegraphics[width=\textwidth]{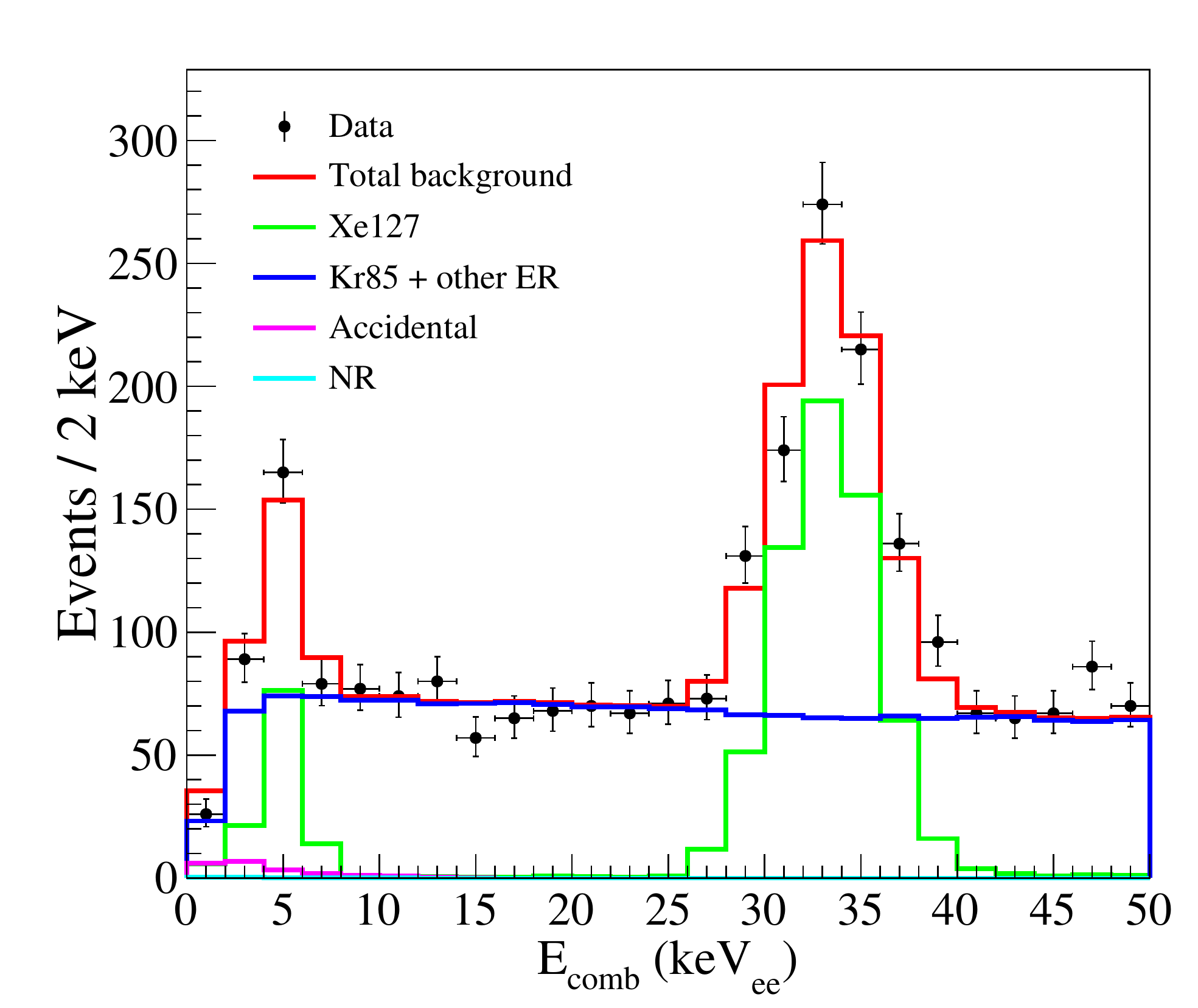} 
  \end{subfigure}
  \begin{subfigure}[b]{0.45\textwidth}
  \includegraphics[width=\textwidth]{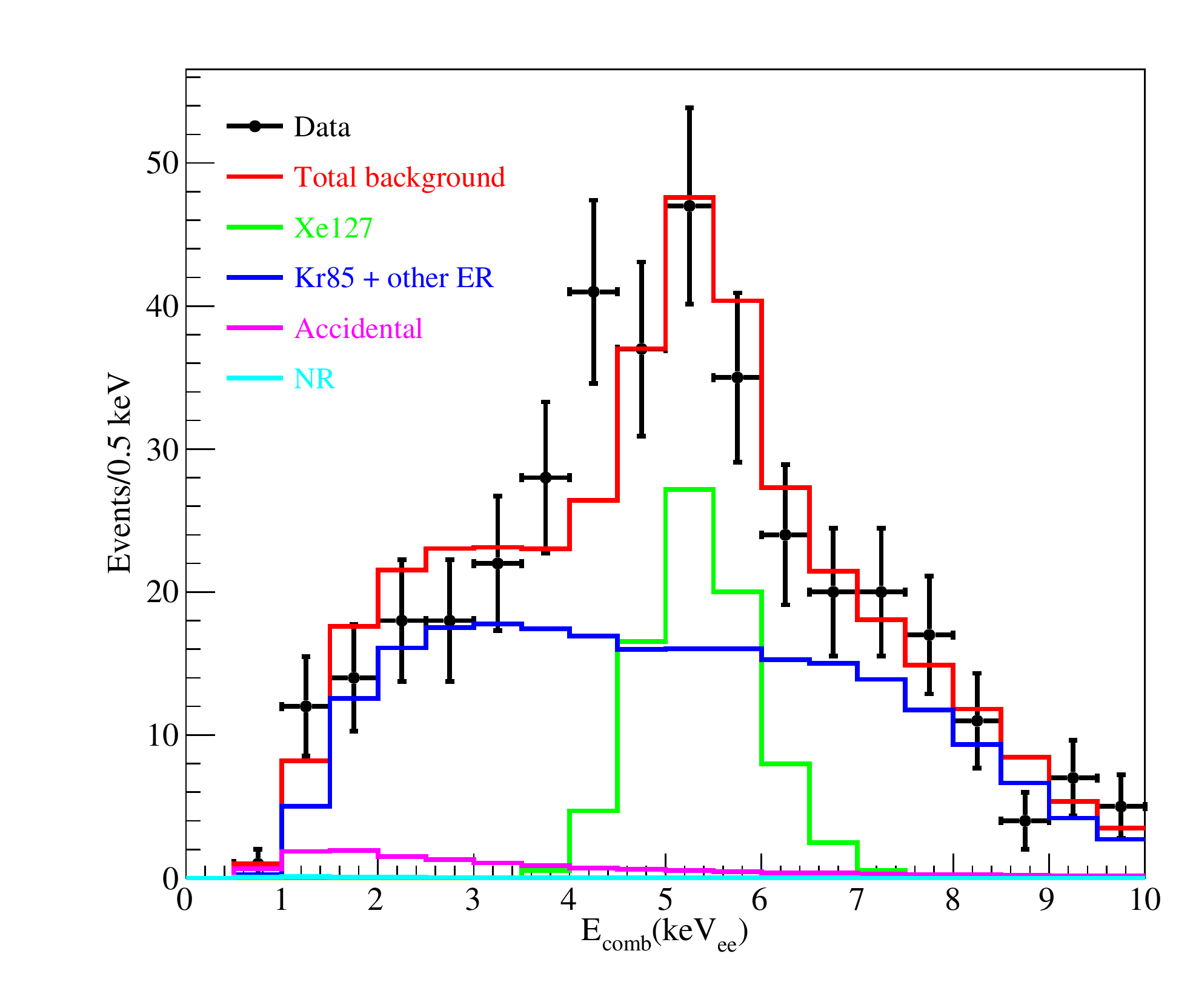} 
  \end{subfigure}
  \caption{Left: combined energy spectrum from 0 to 50 keV in Run 9. Data (black dots) shown include all selection cuts described in Table IV of the main article except that the upper cuts on S1 and S2 are removed. The total background (red) consists of $^{127}$Xe (green), $^{85}$Kr and other ER backgrounds (blue), and neutron background (cyan), all of which are estimated from simulation, as well as accidental background (magenta) estimated from data. When fitting to data, the normalizations of accidental background and NR background were fixed while others were allowed to float. The obtained $^{127}$Xe and $^{85}$Kr rates are consistent with those in Table II of the main article. Right: combined energy spectrum from 0 to 10 keV and individual best fit background components. Data (black dots) shown include all selection cuts described in Table IV of the main article.}
  \label{fig:low_e_bkg}
\end{figure}

\begin{figure}[H]
  \centering
  \includegraphics[width=.45\textwidth]{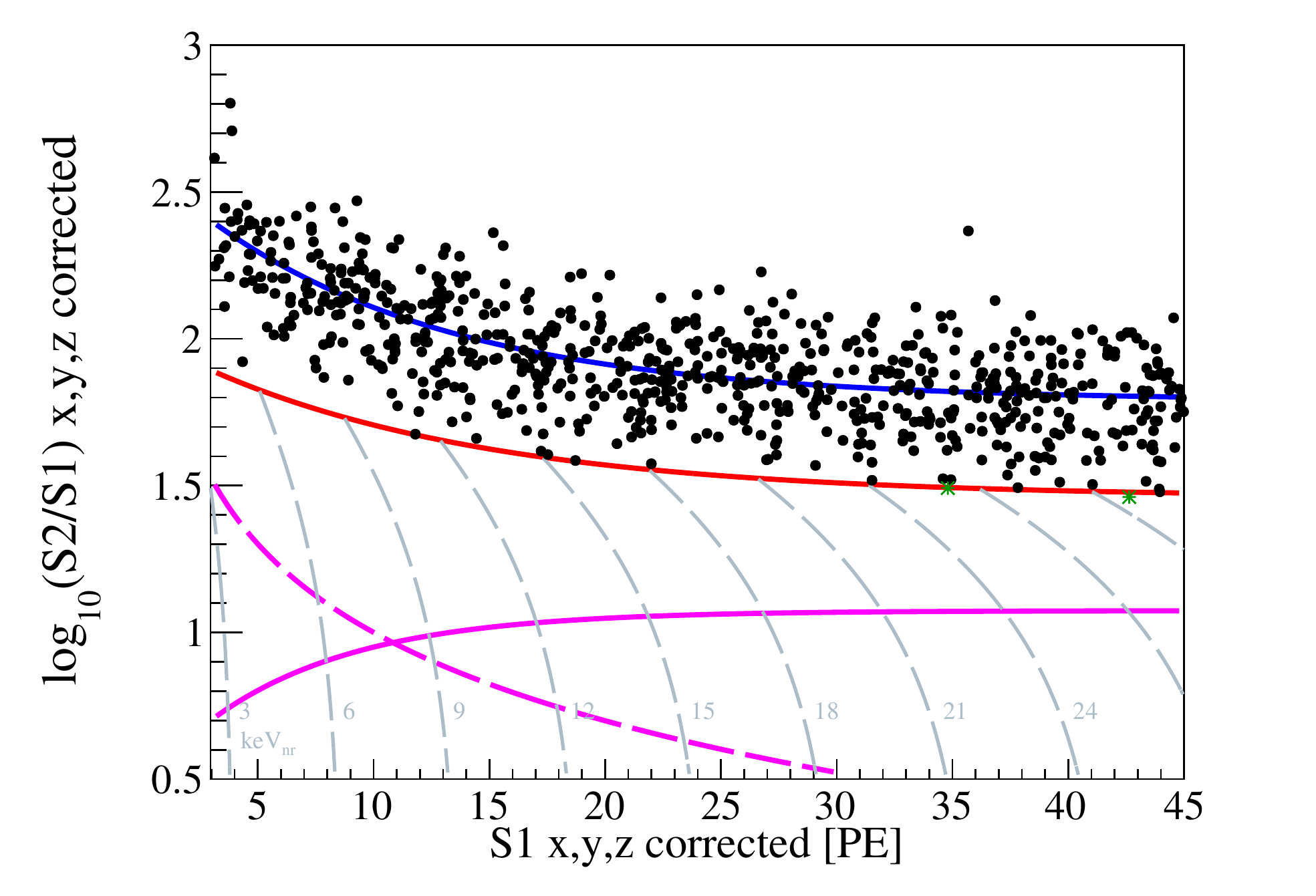}
  \caption{The distribution of log$_{10}(\rm{S2}/\rm{S1})$ versus S1 for DM search data in Run 8 with updated reconstruction and data selection cuts.  The median of the NR calibration band is indicated as the red curve. The dashed magenta curve is the equivalent 100~PE cut on S2. The solid magenta curve is the 99.99\% NR acceptance curve. The gray dashed curves are the equal energy curves with NR energy indicated in the figure. The two data points below the NR median curve are highlighted as green stars.}
  \label{fig:run_8_band}
\end{figure}

\begin{figure}[H]
  \centering
  \includegraphics[width=.45\textwidth]{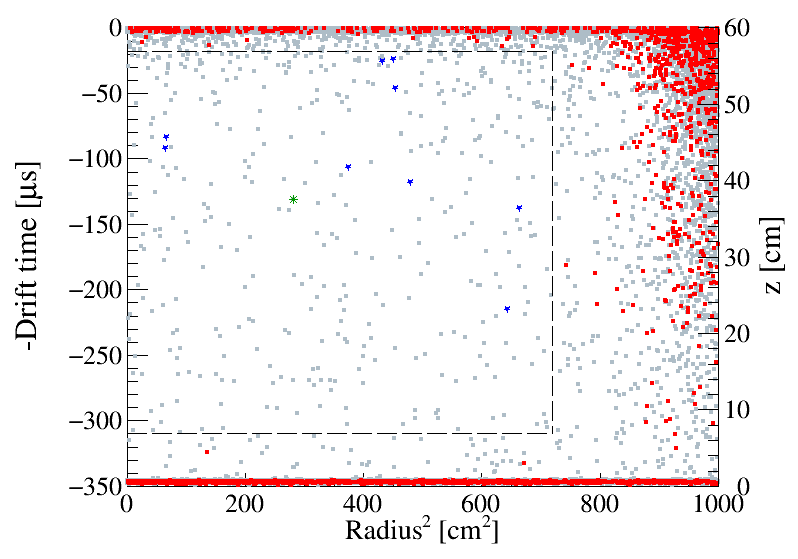}
  \includegraphics[width=.45\textwidth]{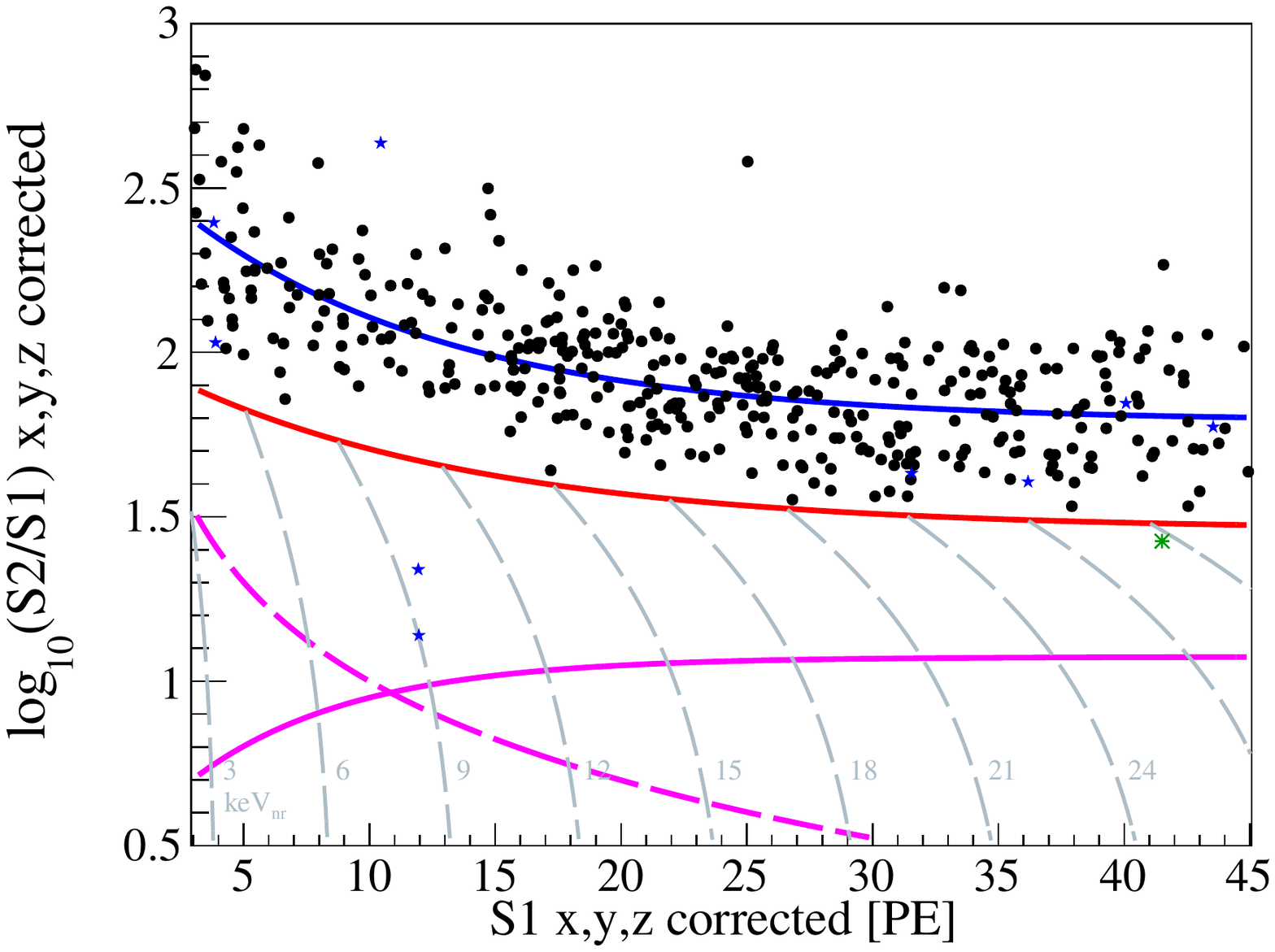}
  \caption{The drift time vs. radius-squared 
    and $\log_{10}$(S2/S1) vs. S1 distributions for candidates in Run 9, with events
    removed by the BDT cut highlighted as the blue stars.}
  \label{fig:candidates_with_bdt_killed}
\end{figure}

\begin{figure}[H]
  \centering
  \includegraphics[width=.45\textwidth]{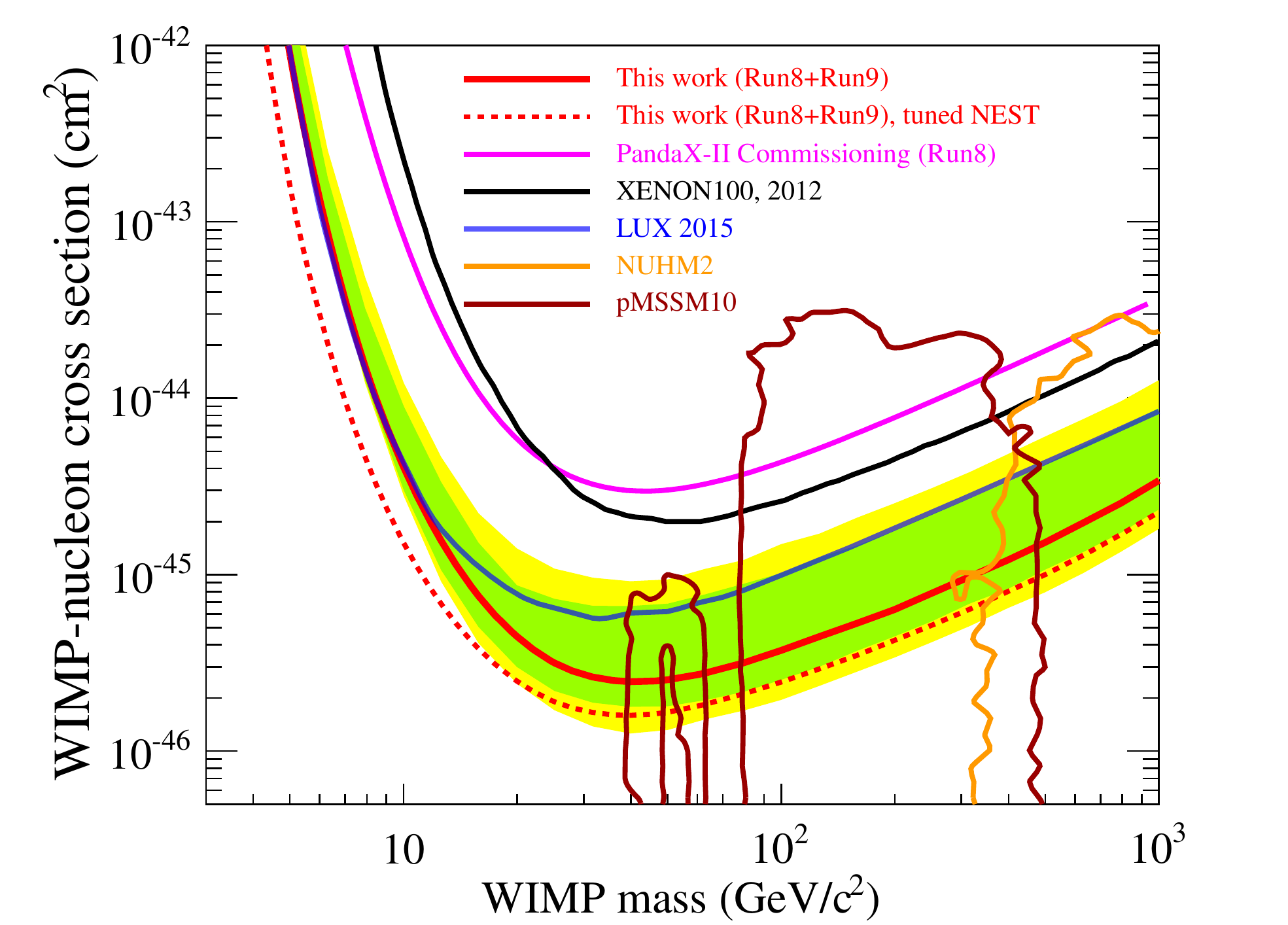}
  \caption{The 90\% C.L. upper limits for the spin-independent isoscalar WIMP-nucleon cross sections from the combination of PandaX-II Runs 8 and 9 data sets, using untuned NEST (solid red line) and tuned NEST (dashed red line) as the model for dark matter 
candidate events, respectively. The 1-$\sigma$ (green) and 2-$\sigma$ (yellow)
sensitivity bands were computed with untuned NEST model. 
Note that the limit from the tuned NEST is more constraining than that was 
presented at (http://idm2016.shef.ac.uk/) 
due to better trained BDT cuts, and is slightly more constraining than 
what LUX presented at the same conference. More cross checking to this NEST tuning is 
needed before we present this as an official result. 
Selected recent world results are plotted for comparison: PandaX-II Run 8 results~\cite{Tan:2016diz} (magenta), XENON100 225 day results~\cite{Aprile:2013doa} (black), and 
LUX 2015 results~\cite{Akerib:2015rjg}(blue). 
Representative supersymmetric model contours (2$\sigma$) after experimental constraints 
from LHC Run 1 (gold and brown) from Ref.~\cite{Bagnaschi:2015eha} 
are overlaid for comparison.
}
  \label{fig:limits_tuned}
\end{figure}

This document contains some detailed figures to support the results presented 
in the main paper, including 
\begin{itemize}
\item Fig.~\ref{fig:e_lifetime}, the evolution of electron lifetime including the data in Runs 8 and 9;
\item Fig.~\ref{fig:doke_plot}, a fit to the anti-correlation between
  S1 and S2 for the ER peaks to extract the PDE and EEE;
\item Fig.~\ref{fig:nest_comparison}, comparison of measured ER light yield and
  charge yield to the prediction by the NEST model.
\item Fig.~\ref{fig:bdt_vs_s1}, 
  the signal (NR) and background (accidental) efficiencies 
  for the BDT cut in S1 and S2 (below-NR-median);
\item Fig.~\ref{fig:bdt_parameters}, 
  comparison of distributions of two example variables between the signal  (NR) and 
  background (accidental) events;
\item Fig.~\ref{fig:nr_comb}, data and MC comparison of AmBe distributions in combined 
  energy, S1 and raw S2, with the MC results from the untuned and tuned NEST overlaid;
\item Fig.~\ref{fig:wall}, drift time vs. reconstructed radius for the 
  wall $^{210}$Po events;
\item Fig.~\ref{fig:low_e_bkg}, measured energy spectra in the dark
  matter data below 50 keV$_{ee}$ and below 10 keV$_{ee}$, and the
  comparison with expected backgrounds;
\item Fig.~\ref{fig:run_8_band}, the distribution of final candidates in 
  log$_{10}(\rm{S2}/\rm{S1})$ vs. S1 from the Run 8 data in this new
  analysis;
\item Fig.~\ref{fig:candidates_with_bdt_killed}, the drift time vs. radius-squared 
  and $\log_{10}$(S2/S1) vs. S1, for the candidates in Run 9, with events
  removed by the BDT cut highlighted as the blue points;
\item Fig.~\ref{fig:limits_tuned}, comparison of limits between
  experiments, and the results from this data using the untuned and  tuned
  NEST model.
\end{itemize}

\end{document}